# A Unified MGF-Based Capacity Analysis of Diversity Combiners over Generalized Fading Channels

Ferkan Yilmaz and Mohamed-Slim Alouini

Electrical Engineering Program, Division of Physical Sciences and Engineering,

King Abdullah University of Science and Technology (KAUST),

Thuwal, Mekkah Province, Saudi Arabia.

Email(s): {ferkan.yilmaz, slim.alouini}@kaust.edu.sa

**Abstract**

Unified exact average capacity results for $L$-branch coherent diversity receivers including equal-gain combining (EGC) and maximal-ratio combining (MRC) are not known. This paper develops a novel generic framework for the capacity analysis of $L$-branch EGC/MRC over generalized fading channels. The framework is used to derive new results for the Gamma shadowed generalized Nakagami-$m$ fading model which can be a suitable model for the fading environments encountered by high frequency (60 GHz and above) communications. The mathematical formalism is illustrated with some selected numerical and simulation results confirming the correctness of our newly proposed framework.

**Index Terms**

Average capacity, diversity, equal-gain combining (EGC), maximal-ratio combining (MRC), correlated channel fading, Gamma shadowed generalized Nakagami-$m$ fading.

## I. INTRODUCTION

Equal gain combining (EGC) is of practical interest in 60 GHz communications because its performance is comparable to that of maximal ratio combining (MRC) but it offers a greater simplicity of implementation (see [1] for an extended discussion on EGC and MRC performance difference). Due to high data-rate and coverage requirements of current, emerging and future high-frequency (60 GHz or above) communication systems, the average capacity (AC) analysis of these two diversity combiners (i.e., EGC and MRC) becomes an important and fundamental issue from both theoretical and practical viewpoints.

In literature, there are several papers dealing with the average symbol error probability (ASEP) analysis of the diversity receivers (see for example [1] and the references therein). Advances over the last decade on the symbol error performance analysis of EGC and MRC diversity receivers in fading channels has accentuated the importance of the moment generating functions (MGF) as a powerful tool for simplifying the analysis of diversity receivers. For example, the following identity has been widely used to simplify the symbol error performance analysis of EGC and MRC diversity receivers in fading channels,

$$\mathrm{erfc}\left(\sqrt{\gamma_{end}}\right) = \frac{2}{\pi}\int_0^\infty \exp\left(-\frac{\gamma_{end}}{\sin^2(\theta)}\right)d\theta, \quad (1)$$

where $\mathrm{erfc}(\cdot)$ is the complementary error function [2, Sec.(6.13)], and where $\gamma_{end}$ is the total signal-to-noise ratio (SNR) at the diversity receiver. On the other hand, and to the best of our knowledge, published papers dealing with the AC analysis of EGC and MRC diversity combiners over fading channels have been scarce when compared to those concerning the ASEP performance [1]. In particular, Bhaskar derived in [3] the average capacity of $L$-branch EGC relying on the Gamma approximation of the sum of mutually independent and identically distributed Rayleigh random variables (RVs). In addition, using an MGF-based approach, Hamdi obtained in [4] a new expression for the average capacity of MRC diversity combiner over arbitrarily correlated Rician fading channels. More recently, Di Renzo *et. al* proposed a new framework in [5] in order to compute the average capacity of MRC diversity combiner over generalized fading channels through the medium of the exponential integral $\mathrm{E_i}$ transform. However, the MGF-based approaches developed in [4], [5] were limited to the capacity of calculation of MRC diversity receivers and are not easily extendible to the computation of the capacity of EGC diversity receivers. In this paper, we show that it is actually possible to express the conditional capacity $\log_2(1 + \gamma_{end})$ in a form similar to (1), which facilitates the development of a new unified MGF-based approach for the calculation of the ergodic capacity in arbitrarily correlated/uncorrelated fading channels. More specifically, we present a unified MGF based average capacity computation not only for the $L$-branch MRC diversity receiver but also for the $L$-branch EGC diversity receiver over a wide variety of fading channels and for an arbitrary number of diversity branches.

The remainder of this paper is organized as follows. In Section II, a unified capacity analysis of diversity receivers over generalized fading channels is introduced and some key results are presented. In Section III, after the introduction of Gamma-shadowed generalized Nakagami-$m$ (GNM) fading channel model, the exact average capacities for the EGC and MRC diversity receivers over Gamma-shadowed GNM fading channels are derived and many special cases are deduced. Numerical examples are then given in Section IV to illustrate the mathematical formalism. Finally, the main results are summarized and some conclusions are drawn in the last section.

## II. AN MGF-BASED CAPACITY ANALYSIS OF DIVERSITY COMBINERS

For EGC and MRC diversity receivers, before the signals on the diversity branches are being summed to form the resultant output, the signals on the diversity branches are first co-phased and then weighted equally in EGC or weighted with the fading envelopes in MRC. The instantaneous SNR $\gamma_{end}$ at the output of the diversity receiver can be generically written as

$$\gamma_{end} = \frac{E_s}{N_0 \sqrt{L^{1-p+q}}} \left( \sum_{\ell=1}^{L} \mathcal{R}_\ell^p \right)^q \tag{2}$$

where the parameters $p \in \{1, 2\}$ and $q \in \{1, 2\}$ are chosen as

$$(p, q) = \begin{cases} (1, 2), & \text{EGC} \\ (2, 1), & \text{MRC} \end{cases}. \tag{3}$$

In (2), $L$ denotes the number of branches and $E_s/N_0$ is the transmitted SNR per symbol, and for $\ell \in \{1, 2, 3, \ldots, L\}$, $\mathcal{R}_\ell$ is the $\ell$th branch fading.

Considering the (instantaneous) Shannon capacity of the diversity receiver (i.e., EGC or MRC) with bandwidth $W$ over fading channels (i.e., $C_{\gamma_{end}} \triangleq W \log_2(1 + \gamma_{end})$), the average ergodic channel capacity defined as $C_{avg} \equiv \mathbb{E}[W \log_2(1 + \gamma_{end})]$, where $\mathbb{E}[\cdot]$ denotes the expectation operator, can be obtained by averaging the instantaneous capacity $C_{\gamma_{end}}$ over the probability density function (PDF) of $\gamma_{end}$, namely

$$C_{avg} = W \int_0^\infty \log_2(1 + \gamma) \, p_{\gamma_{end}}(\gamma) \, d\gamma, \tag{4}$$

where $p_{\gamma_{end}}(\gamma)$ is the PDF of the instantaneous SNR $\gamma_{end}$ (that is, $\gamma_{end}$ is generically defined in (2)). Due to several reasons (e.g., insufficient antenna spacing or coupling among radio frequency (RF) layers), correlation may exist among diversity branches of the receiver. With or without that, the average capacity using (4) involves an $L$-fold integral given by

$$C_{avg} = W \underbrace{\int_0^\infty \int_0^\infty \cdots \int_0^\infty}_{L\text{-fold}} \log_2\left(1 + \frac{E_s}{N_0\sqrt{L^{1-p+q}}}\left(\sum_{\ell=1}^L r_\ell^p\right)^q\right) p_{\mathcal{R}_1,\mathcal{R}_2,\ldots,\mathcal{R}_L}(r_1, r_2, \ldots, r_L) dr_1 dr_2 \ldots dr_L, \quad (5)$$

where $p_{\mathcal{R}_1,\mathcal{R}_2,\ldots,\mathcal{R}_L}(r_1, r_2, \ldots, r_L)$ is the joint multivariate PDF of $\mathcal{R}_1, \mathcal{R}_2, \ldots, \mathcal{R}_L$ fading envelopes. The $L$-fold integration in (5) is tedious and complicated in addition to the fact that it cannot be separated into a product of one dimensional integrals. In addition, it takes a long time to evaluate numerically, especially as the number of branches $L$ increases. Thus, referring to (4), researchers in literature have tried to find the PDF of the instantaneous SNR $\gamma_{end}$ given in (2) in order to find the average capacity. Nevertheless, this technique is often complicated and tedious for generalized fading environment since it involves multiple convolutions / integrals even if the fading envelopes $\mathcal{R}_1, \mathcal{R}_2, \ldots, \mathcal{R}_L$ of the branches are assumed to be independent. Referring to (2), the Jensen's inequality [6, Eq. (12.411)], which is based on concavity of log function such that $\mathbb{E}[\log_2(1 + \gamma_{end})] \leq \log_2(1+\mathbb{E}[\gamma_{end}])$, and fractional moments [6, Eq. (1.511)], which is based on the infinite series of $\log_2(1+\gamma_{end})$ such that $\mathbb{E}[\log_2(1 + \gamma_{end})] = -\sum_{n\geq 1}(-1)^n \mathbb{E}[\gamma_{end}^n]/\log(2^n)$, are commonly used in particular to compute the AC approximately. The other mostly used way to compute the AC hinges upon the inverse Laplace transform (ILT) whereby the PDF of the instantaneous SNR $\gamma_{end}$ can be approximated through the medium of applying the ILT on the MGF $\mathcal{M}_{\gamma_{end}}(s) = \mathbb{E}[\exp(-s\gamma_{end})]$. It is pertinent to say here again that the AC computation of diversity combiners (especially for the MGF of EGC since it is often more difficult than that of MRC due to the fact that it may not be possible to obtain the MGF of EGC receiver in general fading environments) becomes more difficult, problematic, and perplexing as the number of branches (i.e., $L$) increases.

In what follows, we present a new exact and unified MGF-based approach that overcomes the difficulty mentioned above, and offers a generic single integral expression for the average capacity of EGC and MRC diversity combiners over generalized fading channels.

**Theorem 1** (Average Capacity of the Diversity Combiners over Correlated Not-Necessarily Identically Distributed Fading Channels). *The exact average capacity of $L$-branch diversity combiner over mutually not-necessarily independent nor identically distributed fading channels with a bandwidth $W$ is given by*

$$\mathcal{C}_{avg} = \frac{W}{\log(2)} \int_0^\infty \mathrm{C}_q(s) \left[\frac{\partial}{\partial s} \mathcal{M}_{\vec{\mathcal{R}}^p}(\Phi_{p,q} s)\right] ds, \quad (6)$$

*where $p \in \{1,2\}$ and $q \in \{1,2\}$ and are selected based on* (3), *and where the parameter $\Phi_{p,q}$ is defined as $\Phi_{p,q} = \sqrt[q]{E_s/N_0} L^{(p-q-1)/2}$, and where the auxiliary function $C_q(s)$ is given by*

$$C_q(s) = -H_{3,2}^{1,2}\left[\frac{1}{s^q} \left| \begin{array}{c}(1,1),(1,1),(1,q) \\ (1,1),(0,1)\end{array}\right.\right], \tag{7}$$

*where $H_{p,q}^{m,n}[\cdot]$ represents the Fox's H function [7, Eq. (1.1.1)][1,2]. Moreover, $\mathcal{M}_{\vec{\mathcal{R}}^p}(s) \equiv \mathbb{E}[\exp(-s\sum_\ell \mathcal{R}_\ell^p)]$ is the joint MGF for the $p$-exponent of $\vec{\mathcal{R}} \equiv \{\mathcal{R}_1, \mathcal{R}_2, \ldots, \mathcal{R}_L\}$ fading envelopes of the branches.*

*Proof:* See Appendix A. ∎

Note that, in order to find the average capacity, the proposed MGF-based technique in Theorem 1 eliminates the necessity of finding the PDF of the instantaneous SNR $\gamma_{end}$ through the ILT of the joint $p$-exponent MGF $\mathcal{M}_{\vec{\mathcal{R}}^p}(s)$. Shortly, Theorem 1 suggests that one can readily obtain the average capacity of the diversity receiver by using the joint $p$-exponent MGF $\mathcal{M}_{\vec{\mathcal{R}}^p}(s)$. Additionally, the integral in (1) can be readily estimated accurately by employing the Gauss-Chebyshev quadrature (GCQ) formula [11, Eq. (25.4.39)], yielding

$$\mathcal{C}_{avg} \approx \frac{W}{\log(2)} \sum_{n=1}^{N} w_n C_q(s_n) \left\{ \left.\frac{\partial}{\partial s}\mathcal{M}_{\vec{\mathcal{R}}^p}(\Phi_{p,q}s)\right|_{s \to s_n}\right\}, \tag{8}$$

which converges rapidly and steadily, requiring only few terms for an accurate result, where the coefficients $w_n$ and $s_n$ are defined as

$$s_n = \tan\left(\tfrac{\pi}{4}\cos\left(\tfrac{2n-1}{2N}\pi\right) + \tfrac{\pi}{4}\right) \quad \text{and} \quad w_n = \frac{\pi^2 \sin\left(\tfrac{2n-1}{2N}\pi\right)}{4N\cos^2\left(\tfrac{\pi}{4}\cos\left(\tfrac{2n-1}{2N}\pi\right) + \tfrac{\pi}{4}\right)}, \tag{9}$$

respectively, where the truncation index $N$ could be chosen as $N = 50$ to obtain a high level of accuracy. In addition, when there is no correlation between the fading envelopes $\vec{\mathcal{R}} \equiv \{\mathcal{R}_1, \mathcal{R}_2, \ldots, \mathcal{R}_L\}$ for the branches of the diversity receiver, the average capacity is given in the following corollary.

**Corollary 1** (Average Capacity of the Diversity Receiver over Mutually Independent Not Necessarily Identically Distributed Fading Channels). *The exact average capacity of $L$-branch diversity receiver over mutually independent and non-identically distributed fading channels with the bandwidth $W$ is given by*

$$\mathcal{C}_{avg} = \frac{W}{\log(2)} \int_0^\infty C_q(s) \sum_{\ell=1}^{L} \left[\frac{\partial}{\partial s}\mathcal{M}_{\mathcal{R}_\ell^p}(\Phi_{p,q}s)\right] \prod_{\substack{k=1 \\ k \neq \ell}}^{L} \mathcal{M}_{\mathcal{R}_k^p}(\Phi_{p,q}s) ds \tag{10}$$

---

[1] For more information about the Fox's H function, the readers are referred to [7], [8]

[2] Note that the Fox's H function is still not available in standard mathematical software packages such as Mathematica® and Maple™. However, using [9, Eq. (8.3.2/22)], an efficient mathematica implementation of this function is available in [10, Appendix A].

*where, for $\ell \in \{1, 2, \ldots, L\}$, $\mathcal{M}_{\mathcal{R}_\ell^p}(s) \equiv \mathbb{E}\left[\exp\left(-s\mathcal{R}_\ell^p\right)\right]$ is the MGF of the fading $\mathcal{R}_\ell$ that the $\ell$th branch is subjected to.*

*Proof:* When there is no correlation between the fading envelopes $\vec{\mathcal{R}} \equiv \{\mathcal{R}_1, \mathcal{R}_2, \ldots, \mathcal{R}_L\}$, one can readily write $\mathcal{M}_{\vec{\mathcal{R}}^p}(s) = \prod_{\ell=1}^{L} \mathcal{M}_{\mathcal{R}_\ell^p}(s)$, whose derivative with respect to $s$ is given by

$$\frac{\partial}{\partial s}\mathcal{M}_{\vec{\mathcal{R}}^p}(s) = \sum_{\ell=1}^{L}\left[\frac{\partial}{\partial s}\mathcal{M}_{\mathcal{R}_\ell^p}(s)\right]\prod_{\substack{k=1\\k\neq\ell}}^{L}\mathcal{M}_{\mathcal{R}_k^p}(s). \tag{11}$$

Finally, substituting (11) into (6) results in (10), which proves Corollary 1. ∎

Despite the fact that the novel technique represented by Theorem 1 and Corollary 1 are easy to use, the numerical computation of the auxiliary function $C_q(s)$ can also be done using the more familiar Meijer's G function, which is available in standard mathematical software packages such as Mathematica® and Maple™, as shown in the following corollary.

**Corollary 2** (Meijer's G Representation of the Auxiliary Function $C_q(s)$). *The auxiliary function $C_q(s)$ can be given in terms of the more familiar Meijer's G function as follows*

$$C_q(s) = \frac{-1}{\sqrt{q(2\pi)^{1-q}}}G_{q+2,2}^{1,2}\left[\frac{q^q}{2^q}\,\middle|\,\begin{array}{c}1,1,\Xi_{(q)}^{(1)}\\1,0\end{array}\right], \tag{12}$$

*where $\Xi_{(n)}^{(x)} \equiv \frac{x}{n}, \frac{x+1}{n}, \ldots, \frac{x+n-1}{n}$ with $x \in \mathbb{C}$ and $n \in \mathbb{N}$.*

*Proof:* See Appendix B. ∎

Let us consider the special cases ($q=1$ for MRC and $q=2$ for EGC) of the auxiliary function $C_q(s)$ in order to check analytical simplicity and accuracy:

**Special Case 1** (Maximal Ratio Combining). For $L$-branch MRC diversity receiver (i.e., $q=1$), the auxiliary function $C_{MRC}(s) \equiv C_q(s)|_{q\to 1}$ can be obtained as

$$C_{MRC}(s) = -G_{2,1}^{0,2}\left[\frac{1}{s}\,\middle|\,\begin{array}{c}1,1\\0\end{array}\right] \tag{13}$$

by means of applying [9, Eq. (8.2.2/9)] on (12). Utilizing [9, Eq. (8.2.2/14)] and [9, Eq. (8.4.11/1)] together, (13) reduces further to

$$C_{MRC}(s) = \text{Ei}(-s), \tag{14}$$

where $\text{Ei}(\cdot)$ is the exponential integral function [2, Eq. (6.15.2)]. [3]. Then, referring to Theorem 1, the average capacity of the $L$-branch MRC receiver can be given by

$$\mathcal{C}_{avg}^{MRC} = \frac{W}{\log(2)} \int_0^\infty \text{Ei}(-s) \left[ \frac{\partial}{\partial s} \mathcal{M}_{\vec{\mathcal{R}}^2}\left(\frac{E_s}{N_0}s\right) \right] ds \tag{15}$$

which is in perfect agreement with [5, Eq. (7)] when choosing the transmitted power is unit (i.e., $E_s/N_0 = 1$). In addition, when the branches are subjected to mutually independent and non-identical fading distributions, the average capacity $\mathcal{C}_{avg}^{MRC}$ can be also given, referring to Corollary 1, as follows

$$\mathcal{C}_{avg}^{MRC} = \frac{W}{\log(2)} \int_0^\infty \text{Ei}(-s) \sum_{\ell=1}^L \left[ \frac{\partial}{\partial s} \mathcal{M}_{\mathcal{R}_\ell^2}\left(\frac{E_s}{N_0}s\right) \right] \prod_{\substack{k=1 \\ k \neq \ell}}^L \mathcal{M}_{\mathcal{R}_k^2}\left(\frac{E_s}{N_0}s\right) ds. \tag{16}$$

**Special Case 2** (Equal Gain Combining). Note that, referring (7) with $q = 2$, the auxiliary function for $L$-branch EGC diversity receiver, i.e., $\text{C}_{EGC}(s) \equiv \text{C}_q(s)|_{q\to 2}$ can be re-written as

$$\text{C}_{EGC}(s) = -\sqrt{\pi} G_{3,1}^{0,2}\left[\frac{4}{s^2} \left| \begin{array}{c} 1, 1, \frac{1}{2} \\ 0 \end{array} \right. \right], \tag{17}$$

by means of setting $q = 2$ in (12). Eventually, using [9, Eq. (8.4.12/4)], the auxiliary function for $L$-branch EGC diversity receiver $\text{C}_{EGC}(s)$ simplifies to

$$\text{C}_{EGC}(s) = 2\,\text{Ci}(u), \tag{18}$$

where $\text{Ci}(x)$ is the cosine integral function [11, Eq. (5.2.27)][3]. Then, using Theorem 1 with (18), the average capacity of the $L$-branch EGC diversity receiver can be readily expressed as

$$\mathcal{C}_{avg}^{EGC} = \frac{2W}{\log(2)} \int_0^\infty \text{Ci}(u) \left[ \frac{\partial}{\partial s} \mathcal{M}_{\vec{\mathcal{R}}}\left(\sqrt{\frac{E_s}{N_0 L}}s\right) \right] ds \tag{19}$$

in general when the fading $\mathcal{R}_1, \mathcal{R}_2, \ldots, \mathcal{R}_L$ are subjected to are correlated. When the branches are subjected to mutually independent and non-identical fading, the average capacity $\mathcal{C}_{avg}^{EGC}$ can be also given, referring to Corollary 1, as

$$\mathcal{C}_{avg}^{EGC} = \frac{2W}{\log(2)} \int_0^\infty \text{Ci}(u) \sum_{\ell=1}^L \left[ \frac{\partial}{\partial s} \mathcal{M}_{\mathcal{R}_\ell}\left(\sqrt{\frac{E_s}{N_0 L}}s\right) \right] \prod_{\substack{k=1 \\ k \neq \ell}}^L \mathcal{M}_{\mathcal{R}_k}\left(\sqrt{\frac{E_s}{N_0 L}}s\right) ds. \tag{20}$$

In the following section, the model of Gamma shadowed GNM fading channel will be introduced and then the average capacity of both $L$-branch MRC and $L$-branch EGC diversity receivers will be derived for Gamma-shadowed GNM fading channels.

---

[3]Note that both the cosine integral function $\text{Ci}(x) = -\int_x^\infty \cos(t)/t\, dt$ and the exponential integral $\text{Ei}(x) = -\int_{-x}^\infty \exp(-t)/t\, dt$ are implemented as a built-in function in the more popular mathematical software packages such as Mathematica® and Maple™.

## III. AVERAGE CAPACITY OF DIVERSITY COMBINERS OVER GAMMA-SHADOWED GENERALIZED NAKAGAMI-$m$ FADING CHANNELS

As an example for the application of both Theorem 1 and Corollary 1, we assume that Gamma shadowing affects the GNM [10] fading channels, that is, the local mean power of the fading is a Gamma RV. For example, in $60$ GHz non-line-of-sight (NLOS) propagation, standard deviation of shadowing is typically larger than that of propagation at $5$ GHz. It is therefore not a misstep to assume that the local mean power of channel fading is a RV distributed over $(0, \infty)$. Hence, the Gamma-shadowed GNM fading model can be accommodated to pretest the evaluation of different wireless communications in $60$ GHz non-line-of-sight (NLOS) environment. Thus, we first derive the analytical expressions for both PDF and MGF of the fading amplitudes $\vec{\alpha} \equiv \{\alpha_1, \alpha_2, \ldots, \alpha_L\}$ for the branches of $L$-branch EGC in a Gamma-shadowed GNM fading channel. Using these results, we will find the exact average capacity for the EGC over a Gamma-shadowed GNM fading channel, and will give as examples of the simplified expressions for different special cases.

We first derive the analytical expressions for both PDF and MGF of the fading envelopes $\vec{\mathcal{R}} \equiv \{\mathcal{R}_1, \mathcal{R}_2, \ldots, \mathcal{R}_L\}$ for the branches of $L$-branch diversity receiver in a Gamma-shadowed GNM fading channel. Using these results, we will find the exact average capacity for the diversity receiver over a Gamma-shadowed GNM fading channel and enumerate the different special cases.

### A. Gamma-Shadowed Generalized Nakagami-m Fading Channels

Let us consider $L \geq 1$ mutually independent and non-identical GNM RVs $\{\alpha_\ell\}_{\ell=1}^{L}$, each representing the fading amplitude the $L$-branch diversity combiner is subjected to and each having the PDF

$$p_{\alpha_\ell}(\alpha) = \frac{2\xi_\ell}{\Gamma(m_\ell)} \left(\frac{\beta_\ell}{\Omega_\ell}\right)^{\xi_\ell m_\ell} \alpha^{2\xi_\ell m_\ell - 1} e^{-\left(\frac{\beta_\ell}{\Omega_\ell}\right)^{\xi_\ell} \alpha^{2\xi_\ell}}, \quad 0 \geq \alpha \qquad (21)$$

where the parameters $m_\ell \geq 1/2$, $\xi_\ell > 0$ and $\Omega_\ell > 0$ are the fading figure, the shaping parameter and the local mean power of the $\ell$th GNM RV and $\beta_\ell = \Gamma(m_\ell + 1/\xi_\ell)/\Gamma(m_\ell)$. Furthermore, the special or limiting cases of the GNM distribution are well-known in literature as the Rayleigh ($m_\ell = 1, \xi_\ell = 1$), exponential ($m_\ell = 1, \xi_\ell = 1/2$), Half-Normal ($m_\ell = 1/2, \xi_\ell = 1$), Nakagami-$m$ ($\xi_\ell = 1$), Gamma ($\xi_\ell = 1/2$), Weibull ($m_\ell = 1$), lognormal ($m_\ell \to \infty, \xi_\ell \to 0$), and AWGN ($m_\ell \to \infty, \xi_\ell = 1$).

As mentioned before, let the local mean power, $\Omega_\ell$ of the GNM fading amplitude for the $\ell$th branch of the $L$-branch diversity combiner has, due to the shadowing, distribution with the Gamma PDF given by

$$p_{\Omega_\ell}(\Omega) = \frac{1}{\Gamma(m_{s\ell})}\left(\frac{m_{s\ell}}{\Omega_{s\ell}}\right)^{m_{s\ell}} \Omega^{m_{s\ell}-1} \exp\left(-\frac{m_{s\ell}}{\Omega_{s\ell}}\Omega\right), \quad 0 \leq \Omega_{s\ell}, \frac{1}{2} \leq m_{s\ell} \quad (22)$$

where $\Omega_{s\ell}$ is the average power of shadowing in the area of interest, and where $m_{s\ell}$ inversely reflect the shadowing severity such that the severity of the shadowing decreases as the value of $m_{s\ell}$ increases. For example, in the limit case $m_{s\ell} \to \infty$, the distribution of the local mean closes to the Dirac's distribution as $p_{\Omega_\ell}(\Omega) = \delta(\Omega - \Omega_{s\ell})$. Hence, there is no shadowing effects.

Eventually, averaging (21) with respect to $\Omega_\ell$, i.e., $\int_0^\infty p_{\alpha_\ell}(\alpha) p_{\Omega_\ell}(\Omega_\ell) d\Omega_\ell$, then utilizing [7, Theorem 2.9] with [7, Eq. (2.9.4)], we obtain the PDF of the Gamma-shadowed GNM fading as introduced in the following definition.

**Definition 1** (Gamma-Shadowed Generalized Nakagami-*m* RV). *The distribution $\mathcal{R}_\ell$ follows an Gamma-shadowed GNM distribution if the PDF of $\mathcal{R}_\ell$ is given by*

$$p_{\mathcal{R}_\ell}(r) = \frac{2}{\Gamma(m_{s\ell})\Gamma(m_\ell)}\left(\frac{\beta_\ell m_{s\ell}}{\Omega_{s\ell}}\right)^{m_{s\ell}} r^{2m_{s\ell}-1} \Gamma\left(m_\ell - \frac{m_{s\ell}}{\xi_\ell}, 0, \frac{\beta_\ell m_{s\ell}}{\Omega_{s\ell}}r^2, \frac{1}{\xi_\ell}\right) \quad (23)$$

*where $m_\ell$ ($0.5 \leq m_\ell < \infty$) and $\xi_\ell$ ($0 \leq \xi_\ell < \infty$) represent the fading figure (diversity severity / order) and the shaping factor, respectively, while $m_{s\ell}$ ($0.5 \leq m_{s\ell} < \infty$) and $\Omega_{s\ell}$ ($0 \leq \Omega_{s\ell} < \infty$) denote the severity and the average power of shadowing, respectively. Moreover, $\Gamma(\cdot,\cdot,\cdot,\cdot)$ is the extended incomplete Gamma function defined as $\Gamma(\alpha, x, b, \beta) = \int_x^\infty r^{\alpha-1} \exp\left(-r - br^{-\beta}\right) dr$ [12, Eq. (6.2)], where $\alpha$ and $x$ are complex parameters, $\beta > 0$ and $b$ is a complex variable.*

In what follows, the shorthand notation $\mathcal{R} \sim \mathcal{N}_\mathcal{S}(m, \xi, m_s, \Omega_s)$ denotes that $\mathcal{R}$ follows a Gamma-shadowed GNM RV with the fading figure $m$, the shaping parameter $\xi$, the shadowing severity $m_s$ and the shadowing average power $\Omega_s$.

Let us consider some special cases of (23) in order to check validity. In fact, this PDF is a very general shadowed PDF which includes many special cases as explained in the second paragraph of this section. For example, By using [11, Eq. (6.1.47)] with the Mellin-Barnes contour integral representation [7, Eq. (1.1.1)] of (23), the PDF reduces to the PDF of the GNM distribution [10, Eqs. (1) and (2)] for $m_{s\ell} \to \infty$ as expected. Here again, by using [12, Eq. (6.41)], the PDF is reduced into the PDF of Gamma-shadowed Nakagami-*m* distribution [13, Eq. (9)] when the shaping parameter $\xi_\ell = 1$. Furthermore, the special or limiting cases of the Gamma-shadowed GNM distribution are well-known in literature as exponential-shadowed Rayleigh ($m_{s\ell} = 1, m_\ell = 1, \xi_\ell = 1$), K distribution ($m_{s\ell} = 1, \xi_\ell = 1$), generalized-K distribution ($\xi_\ell = 1$), Rayleigh ($m_\ell = 1, \xi_\ell = 1, m_{s\ell} \to \infty$), exponential ($m_\ell = 1, \xi_\ell = 1/2, m_{s\ell} \to \infty$), Half-Normal ($m_\ell = 1/2, \xi_\ell = 1, m_{s\ell} \to \infty$), Nakagami-$m$ ($\xi_\ell = 1, m_{s\ell} \to \infty$), Gamma ($\xi_\ell = 1/2, m_{s\ell} \to \infty$), Weibull ($m_\ell = 1, m_{s\ell} \to \infty$), lognormal ($m_\ell \to \infty, \xi_\ell \to 0, m_{s\ell} \to \infty$), and AWGN ($m_\ell \to \infty, \xi_\ell = 1, m_{s\ell} \to \infty$). Regarding these special and limit cases, the Gamma-shadowed GNM distribution $\mathcal{R} \sim \mathcal{N}_\mathcal{S}(m_\ell, \xi_\ell, m_{s\ell}, \Omega_{s\ell})$ has the advantage of modeling the envelope statistics of most known wireless and optical communication channels. Accordingly, it provides a unified theory as to model the envelope statistics.

Referring to Theorem 1 and Corollary 1, we need to obtain the MGF of the fading envelope $\mathcal{R}_\ell \sim \mathcal{N}_\mathcal{S}(m_\ell, \xi_\ell, m_{s\ell}, \Omega_{s\ell})$, i.e., $\mathcal{M}_{\mathcal{R}_\ell}(s) = \mathbb{E}\left[\exp\left(-s\mathcal{R}_\ell\right)\right]$ for $\Re\{s\} \in \mathbb{R}^+$ in order to find the exact average capacity of EGC, and also we need to obtain the MGF of the fading power $\gamma_\ell \equiv \mathcal{R}_\ell^2$, i.e., $\mathcal{M}_{\mathcal{R}_\ell^2}(s) = \mathbb{E}\left[\exp\left(-s\mathcal{R}_\ell^2\right)\right]$ for $\Re\{s\} \in \mathbb{R}^+$ in order to find the exact average capacity of MRC. As such, in the following theorem (i.e., Theorem 2), these MGF functions are obtained in a unified closed form such that we can readily reduce it to the MGF of the $\ell$th branch of EGC and for that of MRC when the values $p = 1$ and $p = 2$ are selected, respectively.

**Theorem 2** (Unified MGF of Gamma-Shadowed GNM RV). *The unified MGF of the Gamma-shadowed GNM envelope distribution $\mathcal{R}_\ell \sim \mathcal{N}_\mathcal{S}(m_\ell, \xi_\ell, m_{s\ell}, \Omega_{s\ell})$, i.e., $\mathcal{M}_{\mathcal{R}_\ell^p}(s) = \mathbb{E}\left[\exp\left(-s\mathcal{R}_\ell^p\right)\right]$ is given by*

$$\mathcal{M}_{\mathcal{R}_\ell^p}(s) = \frac{4}{\Gamma(m_{s\ell})\Gamma(m_\ell)} \mathrm{H}_{1,2}^{2,1}\left[\left(\frac{\beta_\ell m_{s\ell}}{\Omega_{s\ell}}\right)^p \frac{1}{s^2} \bigg| \begin{matrix} (1,2) \\ (m_{s\ell}, p), (m_\ell, p/\xi_\ell) \end{matrix}\right] \qquad (24)$$

*with the convergence region $\Re\{s\} \in \mathbb{R}^+$.*

*Proof:* Note that the unified MGF $\mathcal{M}_{\mathcal{R}_\ell^p}(s) = \mathbb{E}\left[\exp\left(-s\mathcal{R}_\ell^p\right)\right]$, $\mathcal{R}_\ell \sim \mathcal{N}_\mathcal{S}(m_\ell, \xi_\ell, m_{s\ell}, \Omega_{s\ell})$ can readily be given as $\mathcal{M}_{\mathcal{R}_\ell^p}(s) = \int_0^\infty \exp\left(-sr^p\right) p_{\mathcal{R}_\ell}(r)$, where substituting the Fox's H representation of both extended incomplete Gamma function (i.e., [12, Eq. (6.22)]) and exponential function [7, Eq. (2.9.4)] results in

$$\mathcal{M}_{\mathcal{R}_\ell^p}(s) = \frac{2}{\Gamma(m_{s\ell})\Gamma(m_\ell)} \int_0^\infty \frac{1}{r} \mathrm{H}_{0,1}^{1,0}\left[sr^p \bigg| \begin{matrix} --- \\ (0,1) \end{matrix}\right] \mathrm{H}_{0,2}^{2,0}\left[\frac{\beta_\ell m_{s\ell}}{\Omega r^{-2}} \bigg| \begin{matrix} --- \\ (m_{s\ell}, 1), (m_\ell, 1/\xi_\ell) \end{matrix}\right] dr. \qquad (25)$$

Eventually, applying [7, Theorem 2.3] on (25), one can readily obtain the MGF of the Gamma-shadowed GNM distribution given in (24), which proves Theorem 2. ∎

Now, let us consider some special cases in order to check analytical simplicity and accuracy of (24). When setting the shadowing severity $m_{s\ell} \to \infty$, and applying $\lim_{a\to\infty} \frac{\Gamma(a+b)a^c}{\Gamma(a+c)a^b} \approx 1$, where $|b| \ll a$ and $|c| \ll a$, on the Mellin-Barnes integral representation [7, Eq. (1.1.1)] of (24), the unified MGF is, as expected, reduced to the MGFs of GNM [10, Eq. (2)] and generalized Gamma [14, Eq. (11)] for the values $p = 1$ and $p = 2$, respectively.

Note that the unified MGF given by (24) may lead to some computation difficulty to compute due to the fact that the implementation of the Fox's H function is currently not available in standard mathematical software packages but an Mathematica® implementation of this function is offered by the authors in [10, Appendix A]. As such, it may be useful to represent (24) in terms of Meijer's G function with the aid of [9, Eq. (8.3.2/22)]. More specifically, (26) is the Meijer's G representation of (24) for the rational values of the parameter $\xi$ (that is, we restrict $\xi$ to $\xi = k/l$, where $k$ and $l$ are arbitrary positive integers.), namely

$$\mathcal{M}_{\mathcal{R}_\ell^p}(s) = \frac{\sqrt{l/\pi}(kp)^{m_{s\ell}}(lp)^{m_\ell - 1}}{(2\pi)^{\frac{kp}{2} + \frac{lp}{2} + k - 2}\Gamma(m_\ell)\Gamma(m_{s\ell})} G_{2k,kp+lp}^{kp+lp,2k}\left[\frac{\left(\frac{2k}{s}\right)^{2k}\left(\frac{\beta_\ell m_{s\ell}}{\Omega_{s\ell}}\right)^{kp}}{(kp)^{kp}(lp)^{lp}} \middle| \begin{array}{c} -\Xi_{(2k)}^{(-2k)} \\ \Xi_{(kp)}^{(m_{s\ell})}, \Xi_{(lp)}^{(m_\ell)} \end{array}\right]. \quad (26)$$

It may be useful to notice that the rational representation of $\xi_\ell \in \mathbb{R}^+$ remains essentially unchanged if there does not exist any rational number close enough to $\xi_\ell$, fulfilling the condition $|k/l - \xi_\ell| < \epsilon/l^2$, with $\epsilon$ chosen to be $10^{-2}$. For more accuracy to rationalize $\xi_\ell$, the conditional parameter $\epsilon$ can be chosen much smaller. Nevertheless, the number of coefficients of the Meijer's G function in (26) gets higher as $\epsilon$ gets smaller, so much so that its computation efficiency considerable reduces and its computation latency[4] increases. In consequence, the Fox's H function is preferable in this case since its computation efficiency is much better than that of Meijer's G function in this situation.

## B. Unified Average Capacity of Diversity Combiners

Let us find the derivative of the unified MGF given by either (24) or (26) with respect to $s$ since we need it in order to find the average capacity of $L$-branch diversity combiners operating over Gamma-shadowed GNM fading channels. Referring to the relation with an MGF $\mathcal{M}_{\mathcal{R}_\ell}(s)$ and its derivative, i.e., $\frac{\partial}{\partial s}\mathcal{M}_{\mathcal{R}_\ell}(s) = -\mathbb{E}[\mathcal{R}_\ell \exp(-s\mathcal{R}_\ell)]$, the derivation of the unified MGF for diversity combiners operating over Gamma-shadowed GNM fading channels in give in the following theorem.

**Theorem 3** (Derivative of the Unified MGF of Gamma-Shadowed GNM RV). *The derivative of the unified*

---

[4]The computation latency of Meijer's G function $G_{p,q}^{m,n}[\cdot]$ is primarily addressed by the total number of coefficients $p + q$.

*MGF for the Gamma-shadowed GNM envelope distribution* $\mathcal{R}_\ell \sim \mathcal{N}_\mathcal{S}(m_\ell, \xi_\ell, m_{s\ell}, \Omega_{s\ell})$, *i.e.*, $\frac{\partial}{\partial s}\mathcal{M}_{\mathcal{R}_\ell^p}(s) = -\mathbb{E}\left[\mathcal{R}_\ell^p \exp\left(-s\mathcal{R}_\ell^p\right)\right]$ *is given by*

$$\frac{\partial}{\partial s}\mathcal{M}_{\mathcal{R}_\ell^p}(s) = \frac{2/s}{\Gamma(m_{s\ell})\Gamma(m_\ell)} \mathrm{H}_{2,3}^{3,1}\left[\left(\frac{\beta_\ell m_{s\ell}}{\Omega_{s\ell}}\right)^p \frac{1}{s^2} \middle| \begin{array}{c} (1,2),(0,1) \\ (1,1),(m_{s\ell},p),(m_\ell,p/\xi_\ell) \end{array}\right] \quad (27)$$

*with the convergence region* $\Re\{s\} \in \mathbb{R}^+$.

*Proof:* Using either [7, Eq. (2.2.2)] or [9, Eq. (8.3.2/15)], the proof is obvious. ∎

Again, following the same steps in the derivation of (26), (27) can be represented on the basis of the Meijer's G function for the rational values of the parameter $\xi = k/l$, where $k$ and $l$ are arbitrary positive integers. Accordingly, (27) can be given by

$$\frac{\partial}{\partial s}\mathcal{M}_{\mathcal{R}_\ell^p}(s) = \frac{\sqrt{4l/\pi}(kp)^{m_{s\ell}}(lp)^{m_\ell-1}k}{(2\pi)^{\frac{kp}{2}+\frac{lp}{2}+k-2}\Gamma(m_\ell)\Gamma(m_{s\ell})} \mathrm{G}_{2k+1,kp+lp+1}^{kp+lp+1,2k}\left[\frac{\left(\frac{2k}{s}\right)^{2k}\left(\frac{\beta_\ell m_{s\ell}}{\Omega_{s\ell}}\right)^{kp}}{(kp)^{kp}(lp)^{lp}} \middle| \begin{array}{c} -\Xi_{(2k)}^{(-2k)},0 \\ 1,\Xi_{(kp)}^{(m_{s\ell})},\Xi_{(lp)}^{(m_\ell)} \end{array}\right]. \quad (28)$$

Finally, by employing Corollary 1 with (24) and (27), new exact single-integral expressions for the evaluation of the average capacity $\mathcal{C}_{avg}$ of $L$-branch diversity combiners over Gamma-shadowed GNM fading channels are immediately written as

$$\mathcal{C}_{avg} = \mathcal{G}_L \int_0^\infty \frac{\mathrm{C}_q(s)}{s} \sum_{\ell=1}^L \mathrm{H}_{2,3}^{3,1}\left[\left(\frac{\beta_\ell m_{s\ell}}{\Omega_{s\ell}}\right)^p \frac{1}{\Phi_{p,q}^2 s^2} \middle| \begin{array}{c} (1,2),(0,1) \\ (1,1),(m_{s\ell},p),(m_\ell,p/\xi_\ell) \end{array}\right] \times \\ \prod_{\substack{k=1 \\ k\neq\ell}}^L \mathrm{H}_{1,2}^{2,1}\left[\left(\frac{\beta_k m_{sk}}{\Omega_{sk}}\right)^p \frac{1}{\Phi_{p,q}^2 s^2} \middle| \begin{array}{c} (1,2) \\ (m_{sk},p),(m_k,p/\xi_k) \end{array}\right] ds, \quad (29)$$

where both the coefficient $\Phi_{p,q}$ and the auxiliary function $\mathrm{C}_q(s)$ are defined in Theorem 1. Furthermore, the coefficient $\mathcal{G}_L$ is defined as $\mathcal{G}_L = \frac{2^{L+1}W}{\log(2)}\left[\prod_{\ell=1}^L \Gamma(m_{s\ell})\Gamma(m_\ell)\right]^{-1}$ Additionally, referring to (8) (i.e., by changing the variable of the integration in (29) as $s = \tan(\theta)$ and then using GCQ formula [11, Eq. (25.4.39)]), we specifically get a finite ($N$-terms) sum approximation converging rapidly and steadily and requiring few terms for an accurate result as shown

$$\mathcal{C}_{avg} \approx \mathcal{G}_L \sum_{n=0}^N w_n \frac{\mathrm{C}_q(s_n)}{s_n} \sum_{\ell=1}^L \mathrm{H}_{2,3}^{3,1}\left[\left(\frac{\beta_\ell m_{s\ell}}{\Omega_{s\ell}}\right)^p \frac{1}{\Phi_{p,q}^2 s_n^2} \middle| \begin{array}{c} (1,2),(0,1) \\ (1,1),(m_{s\ell},p),(m_\ell,p/\xi_\ell) \end{array}\right] \times \\ \prod_{\substack{k=1 \\ k\neq\ell}}^L \mathrm{H}_{1,2}^{2,1}\left[\left(\frac{\beta_k m_{sk}}{\Omega_{sk}}\right)^p \frac{1}{\Phi_{p,q}^2 s_n^2} \middle| \begin{array}{c} (1,2) \\ (m_{sk},p),(m_k,p/\xi_k) \end{array}\right], \quad (30)$$

where $w_n$ and $s_n$ are defined in (9). In the sense of both that either special or limit cases of Gamma-shadowed GNM $\mathcal{N}_\mathcal{S}(m_\ell, \xi_\ell, m_{s\ell}, \Omega_{s\ell})$ model are commonly used fading models in the literature, and that the auxiliary function $\mathrm{C}_q(s)$ is a unified expression for diversity combiners (e.g., $q = 1$ for MRC and $q = 2$ for EGC), it is sufficient to show that the average capacity given by (29) is a unified expression not only for commonly used channel fading models but also for the commonly used MRC and EGC diversity combiners. For example, referring (14) (i.e., using the MRC special case of the auxiliary function $\mathrm{C}_q(s)$ given by (7)) and performing algebraic manipulations [7, Eqs. (2.1.1)-(2.1.5)] after choosing $p = 2$ for MRC, it is straight forward to show that, the unified average capacity (29) reduces to the average capacity of $L$-branch MRC diversity combiner over Gamma-shadowed GNM fading channels, namely

$$\mathcal{C}_{avg}^{MRC} = -\frac{\mathcal{G}_L}{2^{L+1}} \int_0^\infty \frac{\mathrm{Ei}(-s)}{s} \sum_{\ell=1}^L \mathrm{H}_{2,3}^{3,1}\!\left[\frac{N_0\beta_\ell m_{s\ell}}{E_s\Omega_{s\ell}s}\,\bigg|\,\begin{array}{c}(1,1),(0,1)\\(1,1),(m_{s\ell},1),(m_\ell,1/\xi_\ell)\end{array}\right] \times$$
$$\prod_{\substack{k=1\\k\neq\ell}}^L \mathrm{H}_{1,2}^{2,1}\!\left[\frac{N_0\beta_k m_{sk}}{E_s\Omega_{sk}s}\,\bigg|\,\begin{array}{c}(1,1)\\(m_{sk},1),(m_k,1/\xi_k)\end{array}\right] ds, \quad (31)$$

Substituting the fading figures $m_\ell \to m$, fading shaping factors $\xi_\ell = 1$, the shadowing severities $m_{s\ell} \to \infty$ and shadowing powers $\Omega_{s\ell} = \Omega$ for all $\ell \in \{1, 2, \ldots, L\}$ in (29), and then using [7, Eqs. (2.1.1), (2.1.2), and (2.9.1)], the average capacity given by (31) reduces to the average capacity over mutually independent and identically distributed Nakagami-$m$ fading channels,

$$\mathcal{C}_{avg}^{MRC} = -\frac{WL}{\log(2)\Gamma^L(m)} \int_0^\infty \frac{\mathrm{Ei}(-s)}{s} \mathrm{G}_{2,2}^{2,1}\!\left[\frac{N_0 m}{E_s\Omega s}\,\bigg|\,\begin{array}{c}1,0\\1,m\end{array}\right] \mathrm{G}_{1,1}^{1,1}\!\left[\frac{N_0 m}{E_s\Omega s}\,\bigg|\,\begin{array}{c}1\\m\end{array}\right]^{L-1} ds. \quad (32)$$

Subsequently, note that we have $\mathrm{G}_{1,1}^{1,1}[u|\,{}^1_a] = u^a(1+u)^{-a}\Gamma(a)$ [9, Eq. (8.4.2/5)] and $\mathrm{G}_{2,2}^{2,1}[u|\,{}^{1,0}_{1,a}] = -u^a(1+u)^{-a-1}\Gamma(a)$ [9, Eq. (8.4.49/14)]. The average capacity of MRC diversity can be then attained in closed-form through the instrumentality of the Ei-transform equality $\int_0^\infty \mathrm{Ei}(-u)(1+au)^{-b}du = -1/\Gamma(b)\mathrm{G}_{3,2}^{1,3}[a|\,{}^{0,0,1-b}_{1,-1}]$, where $a, b \in \mathbb{R}^+$, that is,

$$\mathcal{C}_{avg}^{MRC} = \frac{WL}{\log(2)\Gamma(mL+1)} \mathrm{G}_{3,2}^{1,3}\!\left[\frac{\bar{\gamma}}{m}\,\bigg|\,\begin{array}{c}0,0,-mL\\0,-1\end{array}\right], \quad (33)$$

where $\bar{\gamma} \triangleq E_s\Omega/N_0$ is the average SNR recovered by one branch of the MRC diversity combiner. Note that (33) represents an alternative compact closed-form expression (that is not limited to integer values of the fading figure $m$) to the result presented in either [15, Eqs. (19) and (20)] or [16, Eqs. (23) and (24)].

In addition, by choosing $p = 1$ and $q = 2$ for EGC, the auxiliary function $\mathrm{C}_q(s)$ simplifies into (18) and the unified average capacity given by (29) reduces to the average capacity of $L$-branch EGC diversity combiner over Gamma-shadowed GNM fading channels, that is,

$$\mathcal{C}_{avg}^{EGC} = \frac{\mathcal{G}_L}{\sqrt{\pi^L}} \int_0^\infty \frac{\text{Ci}(s)}{s} \sum_{\ell=1}^L \text{H}_{2,2}^{2,2}\left[\frac{4LN_0\beta_\ell m_{s\ell}}{E_s\Omega_{s\ell}s^2} \middle| \begin{matrix}(0,1),(\frac{1}{2},1)\\(m_{s\ell},1),(m_\ell,\frac{1}{\xi_\ell})\end{matrix}\right] \times$$
$$\prod_{\substack{k=1\\k\neq\ell}}^L \text{H}_{2,2}^{2,2}\left[\frac{4LN_0\beta_k m_{sk}}{E_s\Omega_{sk}s^2} \middle| \begin{matrix}(1,1),(\frac{1}{2},1)\\(m_{sk},1),(m_k,\frac{1}{\xi_k})\end{matrix}\right] ds, \quad (34)$$

which is obtained after using [7, Eqs. (2.1.1)-(2.1.5)]. Eventually, substituting the fading figures $m_\ell \to m$, fading shaping factors $\xi_\ell = 1$, the shadowing severities $m_{s\ell} = m_s$ and shadowing powers $\Omega_{s\ell} = \Omega$ for all $\ell \in \{1, 2, \ldots, L\}$ in (34), and then using [7, Eqs. (2.1.1), (2.1.2), and (2.9.1)], the average capacity given by (34) reduces to the average capacity over mutually independent and identically distributed generalized-K fading channels as

$$\mathcal{C}_{avg}^{EGC} = \frac{L\mathcal{G}_L}{\sqrt{\pi^L}} \int_0^\infty \frac{\text{Ci}(s)}{s} G_{2,2}^{2,2}\left[\frac{4LN_0 mm_s}{E_s\Omega_s s^2} \middle| \begin{matrix}0,\frac{1}{2}\\m_s,m\end{matrix}\right] G_{2,2}^{2,2}\left[\frac{4LN_0 mm_s}{E_s\Omega_s s^2} \middle| \begin{matrix}1,\frac{1}{2}\\m_s,m\end{matrix}\right]^{L-1} ds, \quad (35)$$

which can be readily computed by means of GCQ rule as seen in (8).

It might be useful to notice that Tables I-III offer simplified expressions for the unified MGF and its derivative, for the variety of commonly used generalized fading channels in order to facilitate for the readers the use of our average capacity results for both MRC and EGC.

## IV. Numerical Results

In this section, we provide some selected numerical results for the previous example, illustrating the average capacity of $L$-branch diversity receiver over Gamma-shadowed GNM fading channels. As seen in all figures (from Fig. 1 to Fig. 4), MRC gives better capacity/performance than EGC as expected, however it is a complex technique since it requires the envelope estimation of the channel fading. In addition, the minimum difference between their performances is obtained by two-branch combining.

In Fig. 1, the average capacity of diversity receiver is depicted with respect to SNR (i.e., $E_s/N_0$) for difference number of branches with Gamma shadowed GNM fading parameters $\forall \ell \in \{1, 2, \ldots, L\}$, $m_\ell = 2, \xi_\ell = 2, m_{s\ell} = 3$ and $\Omega_{s\ell} = 1$. This figure also displays the capacity per unit bandwidth (i.e., $W = 1$). Increasing the number of branches, i.e., $L \gg 2$, the average capacity increases but note that, regarding the relation among diversity gain and number of antennas, the diversity gain obtained by adding an antenna/branch decreases as the total number of antennas/branchs $L$ increases. Note again that selected numerical and simulation results are in perfect agreement.

Amount of fading (AF) is another important statistical characteristic of fading channels, particularly in the context of applying diversity techniques in the transmission of the signals from transmitter to the receiver such as relay technologies. Shortly, this AF is associated with the fading figure / diversity order $m_\ell$ of the PDF given by (23), and for $\mathcal{R}_\ell \sim \mathcal{N}_\mathcal{S}(m_\ell, \xi_\ell, m_{s\ell}, \Omega_{s\ell})$, it is defined as $m_\ell \equiv \mathbb{E}^2\left[\mathcal{R}_\ell^{\xi_\ell}\right] / \mathrm{Var}\left[\mathcal{R}_\ell^{\xi_\ell}\right]$, where $\mathrm{Var}\left[\cdot\right]$ and $\mathbb{E}\left[\cdot\right]$ are the variance and the expectation operators, respectively. As seen in Fig. 2, note that the large diversity gain is obtained by increasing fading figure / diversity order from $0.5$ to $2.0$. For example, a relay between transmitter and receiver or adding one more antenna to the transmitter may increase the diversity order from $1.0$ to almost $2.0$. For $m_\ell \gg 2$, increasing the fading figure gradually and linearly increases the average capacity. In other words, increasing the number of relays or the number of the antennas at the transmitter more than 2 gradually and linearly increases the average capacity. Finally, note again that numerical and simulation results are in perfect agreement.

When the signal recovered by the $\ell$th branch of $L$-branch diversity receiver from the wireless channel ($\ell \in \{1, 2, \ldots, L\}$) is composed of clusters of a multipath wave, each of which propagates in non-homogeneous environment[5] such that they possess similar delay times and with the delay-time spreads of different clusters and their phases are independent [17]–[19]. In this case, the envelope of the received signal, i.e., $\mathcal{R}_\ell$ is considered as a non-linear function of multipath components. More specifically, let $\mathcal{X}_\ell$ and $\mathcal{Y}_\ell$ be the in-phase and quadrature Gaussian elements of the signal recovered from the $\ell$th branch of the $L$-branch diversity receiver. Then, the envelope is represented as $|\mathcal{X}_\ell + \mathrm{i}\mathcal{Y}_\ell|^{\frac{1}{\xi_\ell}}$, where $\mathrm{i} = \sqrt{-1}$ is the imaginer number and where $\xi_\ell$ is the shaping factor. Shortly, the shaping factor is sometimes not a sufficiently qualified parameter to comprehend and contemplate the fading conditions in some wireless communication applications. As such, for the PDF $p_{\mathcal{R}_\ell}(r)$ of the fading envelope $\mathcal{R}_\ell$, the tail properties, i.e., both $\lim_{r \to \infty}\left[p_{\mathcal{R}_\ell}(r)\right]$ and $\lim_{r \to \infty}\left[\partial p_{\mathcal{R}_\ell}(r)/\partial r\right]$ are changed by the shaping factor $\xi_\ell$. As seen in Fig. 3, the average capacity goes to zero when the shaping factor $\xi$ goes to zero ($\forall \ell \in \{1, 2, \ldots, L\}$, $\xi_\ell = \xi$) because, for $0 < \xi \ll 1$, the tail properties approach to zero very fast with respect to the possible envelope values $r \in [0, \infty)$, i.e., $\lim_{r \to \infty}\left[p_{\mathcal{R}_\ell}(r)\right] = 0^+$ and $\lim_{r \to \infty}\left[\partial p_{\mathcal{R}_\ell}(r)/\partial r\right] = 0^-$. Also note that, for the higher values of shaping factor $\xi \gg 1$, the average capacity very gradually and linearly increases as seen in Fig. 3 as the shaping factor $\xi$ increases.

As mentioned at the beginning of the previous section, the link quality is affected by variation of the local mean power due to the shadowing caused by moving obstacles, scatters and reflectors between transmitter and the receiver. The intensity of shadowing on the branches of $L$-branch diversity receiver is characterized by the shadowing factors $m_{s\ell} \in [0.5, \infty)$ for the branches $\ell \in \{1, 2, \ldots, L\}$. In Fig. 4, the average capacity is depicted with respect to shadowing factor $m_s$ (i.e., $\forall \ell \in \{1, 2, \ldots, L\}$, $m_{s\ell} = m_s$). Note that, as seen in Fig. 4, the average capacity does not change as the shadowing factor $m_s$ goes to infinity (i.e., $m_s \to \infty$) since the variation of the local mean power diminishes as $m_s$ increases.

---

[5]Note that non-homogeneous wireless communications environment is very common in high frequencies such as 60 GHz or above due to the fact that the wave-length is very small when it is compared with the non-homogeneous (singular) environment.

## V. CONCLUSION

In this paper, we presented a unified framework to compute the average capacity of diversity combining schemes (i.e., EGC and MRC) over fading channels. We also proposed a versatile fading model, which we term Gamma-shadowed GNM fading, in order to characterize the fading environment in high frequencies such as $60$ GHz and above. Additionally, we derived novel closed-form expressions for the moment generating function (MGF) of both Gamma shadowed GNM fading and its special cases. Some selected simulations have been carried out for different scenarios of fading environment in order to verify the accuracy of the presented framework. Numerical and simulation results are in perfect agreement.

## ACKNOWLEDGMENTS

This work was supported by King Abdullah University of Science and Technology (KAUST).

## APPENDIX A
## PROOF FOR THEOREM 1

Note that, for $q \in \{1, 2\}$ (i.e., $q = 1$ for MRC combining, and $q = 2$ for EGC combining), using the derivation equality $\partial \log (1 + yX^q) / \partial y = R^q / (1 + yR^q)$, we can readily show that

$$\frac{1}{X} \log (1 + X^q) = \int_0^1 \frac{1}{u} \left[ \frac{X^{q-1}}{\frac{1}{u} + X^q} \right] du \qquad (A.1)$$

for $n \in \mathbb{R}^+$. Using the equality $\int_0^\infty z^{\beta-1} \exp(-sz) E_{\alpha,\beta}(-yz^\alpha) \, dz = s^{\alpha-\beta}/(s^\alpha + y)$ [20, Eq. (5.2.3)], where $E_{\alpha,\beta}(\cdot)$ is the Mittag-Leffler function [21, Eq. (1)], we get

$$\frac{1}{X} \log (1 + X^q) = \int_0^\infty \exp(-sX) \left[ \int_0^1 \frac{1}{u} E_{q,1}\left(-\frac{s^q}{u}\right) du \right] ds. \qquad (A.2)$$

Upon substituting $\frac{\partial}{\partial s}\exp(-sX) = -X\exp(-sX)$ into (A.3), and then applying the well-known Leibnitz rule [11], it is easily shown that $\log(1+X^q)$ can be expressed as

$$\log(1+X^q) = -\int_0^\infty \left[\frac{\partial}{\partial s}\exp(-sX)\right]\left[\int_0^1 \frac{1}{u}E_{q,1}\left(-\frac{s^q}{u}\right)du\right]ds. \tag{A.3}$$

After substituting the Mellin-Barnes representation of the Mittag-Leffler function, i.e., $E_{\alpha,\beta}(z) = \frac{1}{2\pi i}\oint_C \Gamma(p)\Gamma(1-p)$ [21, Eq. (3)] and performing algebraic manipulations, (A.3) can immediately be expressed as

$$\log(1+X^q) = -\int_0^\infty \left[\frac{\partial}{\partial s}\exp(-sX)\right] H_{3,2}^{1,2}\left[\frac{1}{s^q}\,\bigg|\,\begin{array}{c}(1,1),(1,1),(1,q)\\(1,1),(0,1)\end{array}\right]ds, \tag{A.4}$$

by favor of the Mellin-Barnes representation of Fox's H function [7, Eq. (1.1.1)]. Eventually, substituting (A.4) into (5) and using some algebraic manipulations, the average capacity of linear diversity receivers (EGC and MRC) can be readily given as in (6), which proves Theorem 1.

## APPENDIX B
## PROOF FOR COROLLARY 2

Note that, by means of [7, Eq. (1.1.1)], the auxiliary function $C_q(s)$ given in (7) can be represented in terms of Mellin-Barnes integral as

$$C_q(s) = -\frac{1}{2\pi i}\oint_C \frac{\Gamma(1+z)\Gamma(-z)\Gamma(-z)}{\Gamma(1-z)\Gamma(1+qz)}s^{qz}dz, \tag{B.1}$$

with the convergence region $\Re\{C\} \in (-1,0)$, where i is the imaginary number (i.e., $i = \sqrt{-1}$). Then, substituting Gauss' multiplication formula $\Gamma(nz) = (2\pi)^{\frac{1}{2}(1-n)}n^{nz-\frac{1}{2}}\prod_{k=1}^n \Gamma\left(z+\frac{k-1}{n}\right)$ [11, Eq. (6.1.20)] into (B.1) and using some algebraic manipulations, we get

$$C_q(s) = \frac{-1}{\sqrt{q(2\pi)^{1-q}}}\left\{\frac{1}{2\pi i}\oint_C \frac{\Gamma(1+z)\Gamma(-z)\Gamma(-z)}{\Gamma(1-z)\prod_{k=1}^q \Gamma\left(\frac{k}{q}+z\right)}\left(\frac{q^q}{s^q}\right)^{-z}dz\right\}. \tag{B.2}$$

Finally, applying [7, Eq. (2.9.1)] on the parenthesis part of (B.2), the auxiliary function $C_q(s)$ can be derived as in (12), which proves Corollary 2.

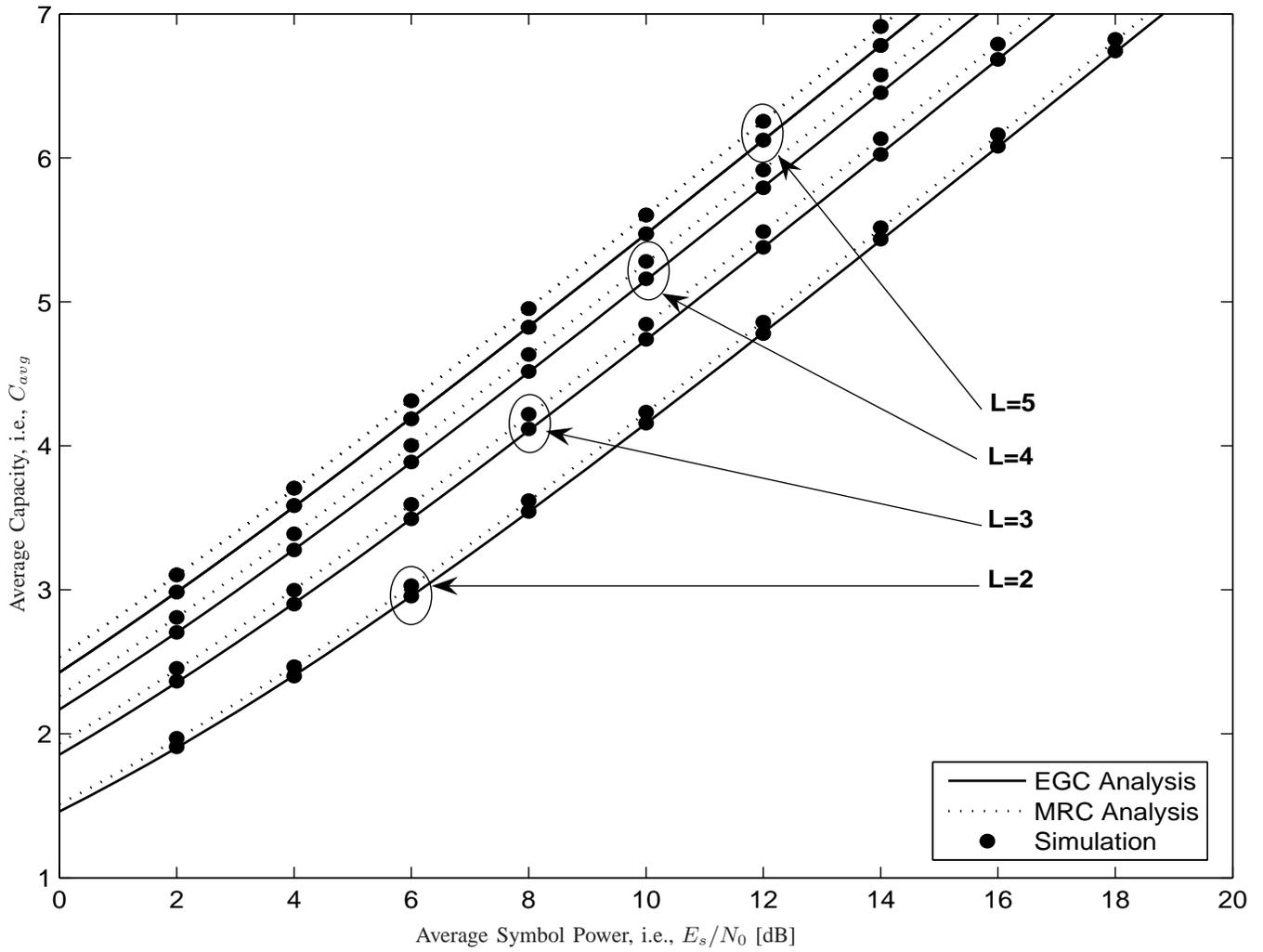

Fig. 1. Average capacity versus the average power for different number of branches over Gamma-shadowed GNM fading channels ($\forall \ell \in \{1, 2, \ldots, L\}$, $m_\ell = 2, \xi_\ell = 2, m_{s\ell} = 3$ and $\Omega_{s\ell} = 1$). The number of samples for the simulation is chosen as $N = 10000$.

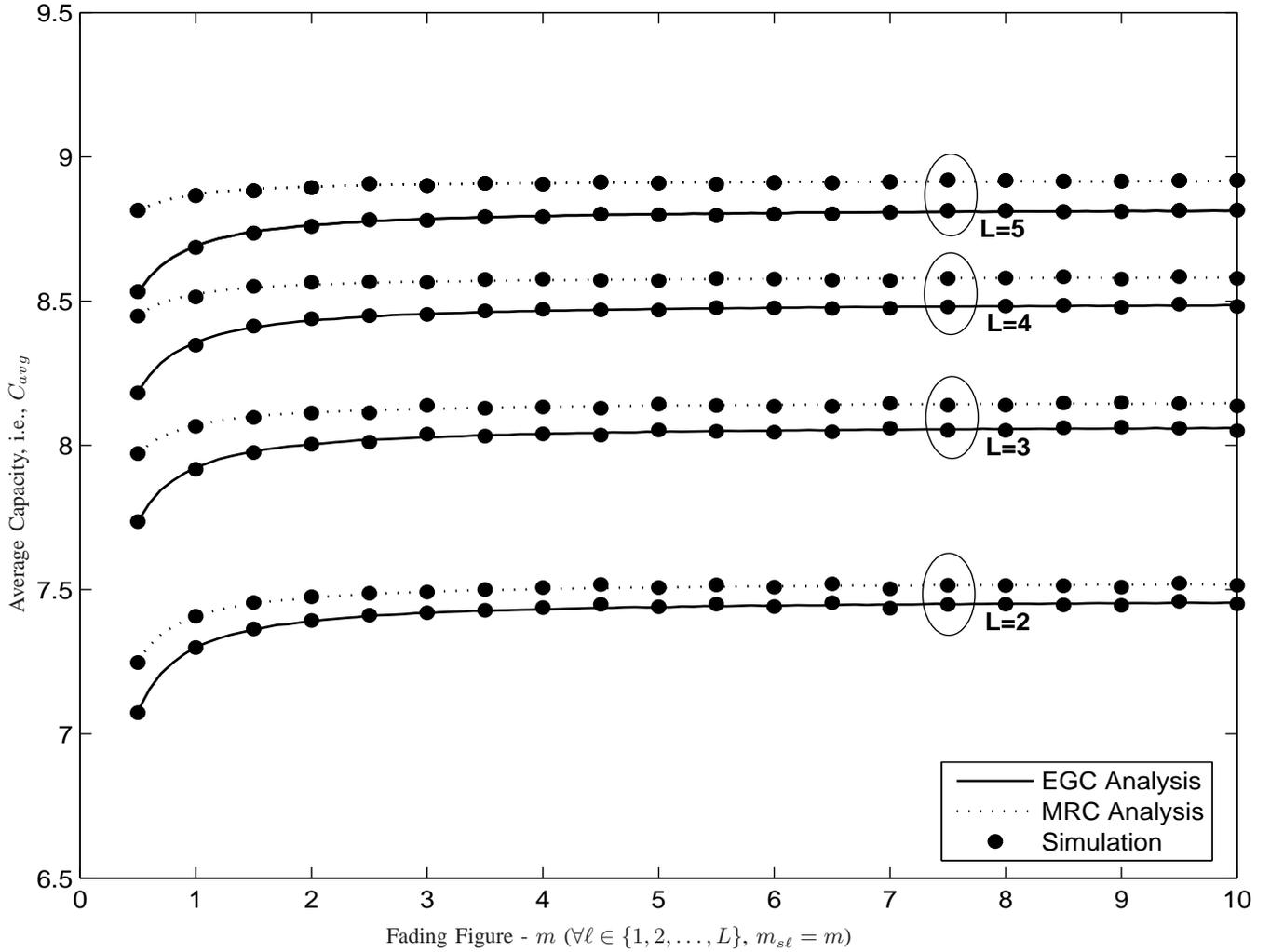

Fig. 2. Average capacity versus the channel fading figure for different number of branches over Gamma-shadowed GNM fading channels ($\forall \ell \in \{1, 2, \ldots, L\}$, $\xi_\ell = 2, m_{s\ell} = 3$ and $\Omega_{s\ell} = 1$). The number of samples for the simulation is chosen as $N = 10000$.

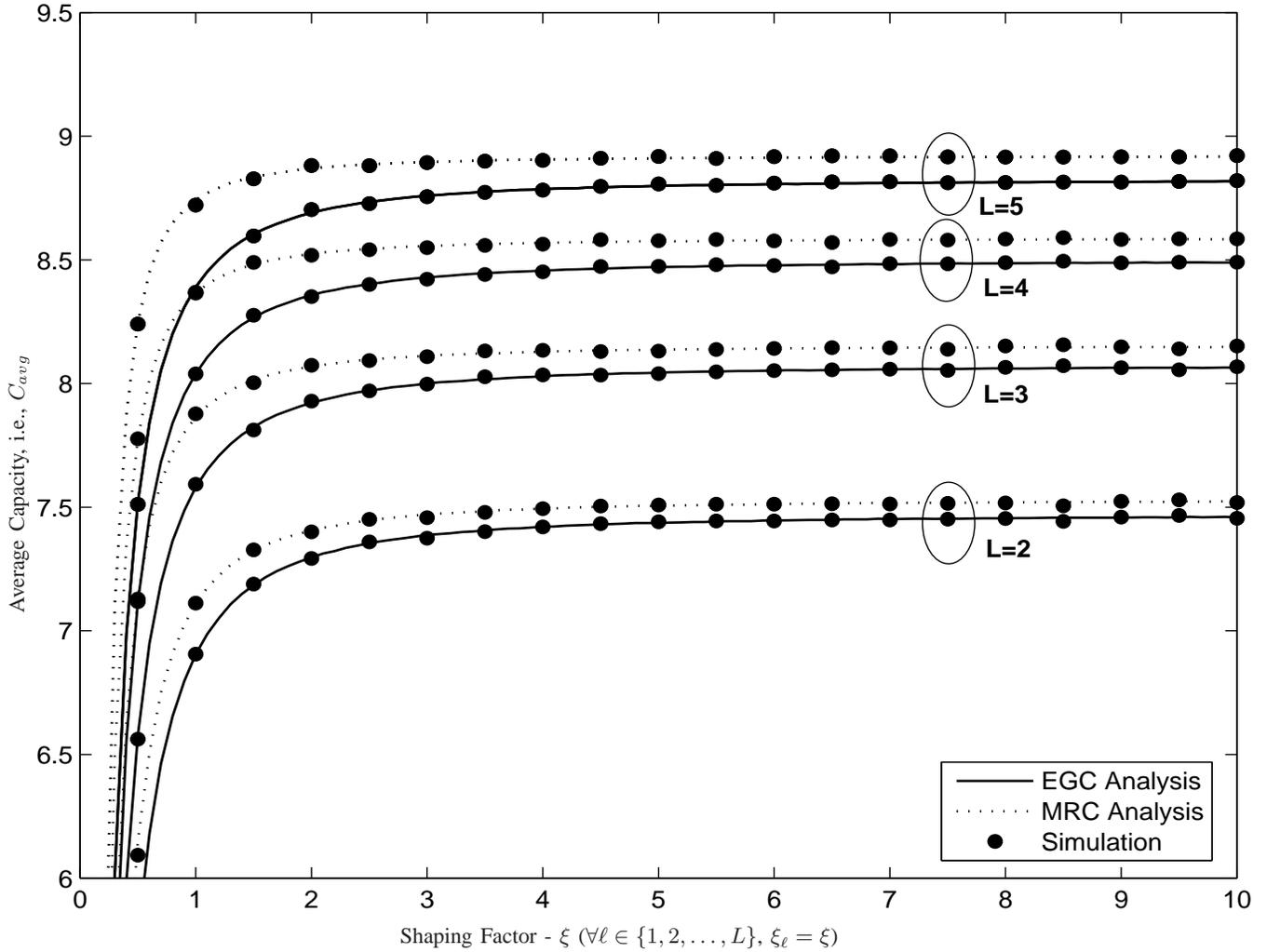

Fig. 3. Average capacity versus the shaping factor for different number of branches over Gamma-shadowed GNM fading channels ($\forall \ell \in \{1, 2, \ldots, L\}$, $m_\ell = 2, m_{s\ell} = 3$ and $\Omega_{s\ell} = 1$). The number of samples for the simulation is chosen as $N = 10000$.

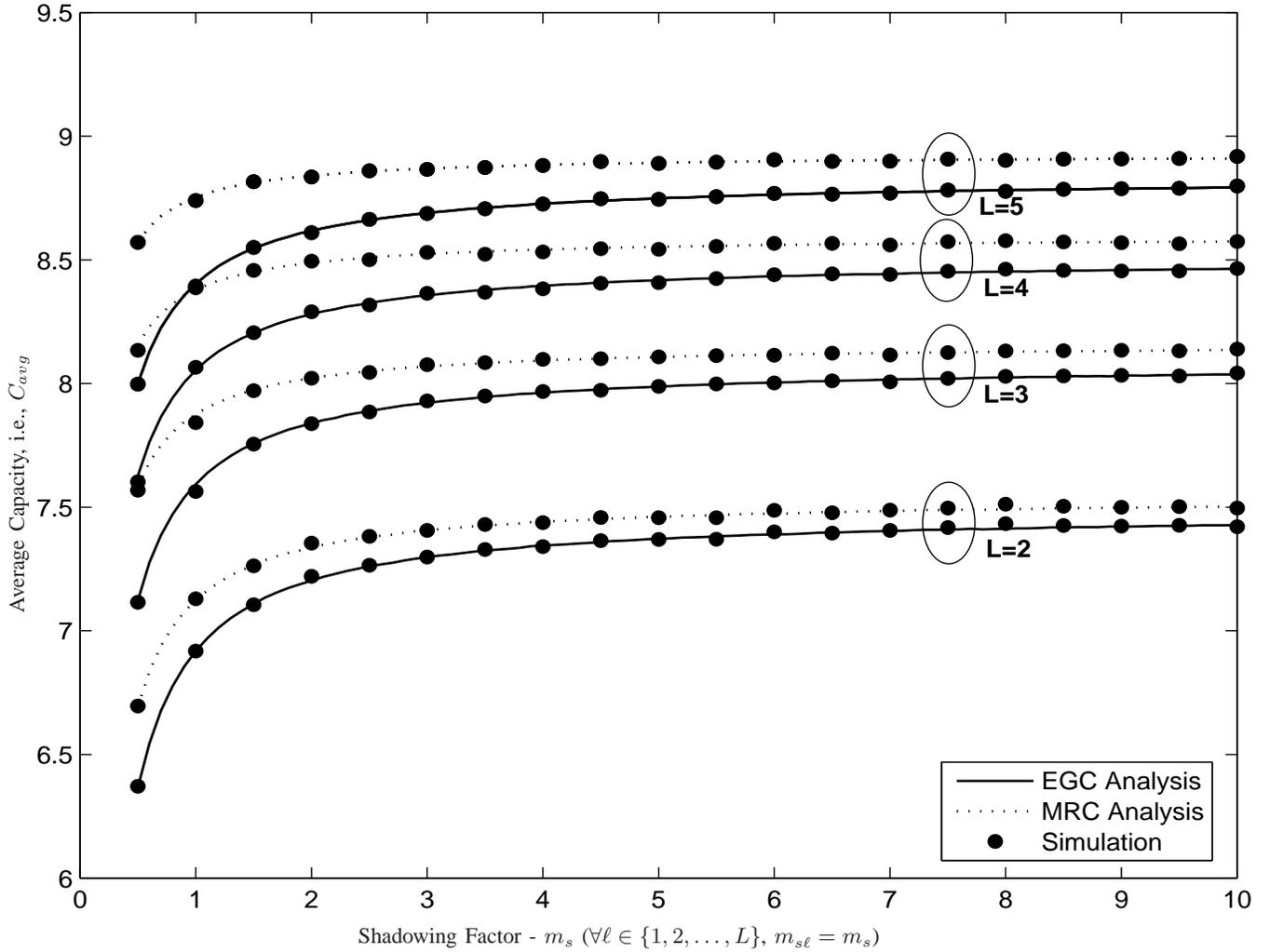

Fig. 4. Average capacity versus the shadowing factor for different number of branches over Gamma-shadowed GNM fading channels ($\forall \ell \in \{1, 2, \ldots, L\}$, $m_\ell = 2, \xi_\ell = 2, m_{s\ell} = 3$ and $\Omega_{s\ell} = 1$). The number of samples for the simulation is chosen as $N = 10000$.

| Envelope Distribution, i.e., $p_{\mathcal{R}_\ell}(r)$ | Unified MGF $\mathcal{M}_{\mathcal{R}_\ell^p}(s)$ and its derivative $\frac{\partial}{\partial s}\mathcal{M}_{\mathcal{R}_\ell^p}(s)$, where the exponent $p \in \{1, 2\}$ |
|---|---|
| **One-Sided Gaussian** [1, Sec. 2.2.1.4] $$p_{\mathcal{R}_\ell}(r) = \sqrt{\frac{2}{\pi\Omega_\ell}} \exp\left(-\frac{r^2}{2\Omega_\ell}\right)$$ defined over $r \in \mathbb{R}^+$, and where $\Omega_\ell$ is the average power (i.e., $\Omega_\ell \geq 0$). Note that one-sided Gaussian fading coincides with the worst-case fading or equivalently, the largest amount of fading (AoF) for all Gaussian-based fading distributions. | $\mathcal{M}_{\mathcal{R}_\ell^p}(s) = \frac{2}{\sqrt{\pi}} H_{1,1}^{1,1}\left[\frac{1}{s^2}\left(\frac{1}{2\Omega_\ell p}\right)^p \bigg| \begin{array}{c}(1,2)\\(\frac{1}{2},p)\end{array}\right] = \frac{2}{\sqrt{(2\pi)^{p+1}}} G_{2,p}^{p,2}\left[\frac{4}{s^2}\left(\frac{1}{2\Omega_\ell p}\right)^p \bigg| \begin{array}{c}\frac{1}{2},1\\\Xi_{(p)}^{(\frac{1}{2})}\end{array}\right],$ $\frac{\partial}{\partial s}\mathcal{M}_{\mathcal{R}_\ell^p}(s) = \frac{4}{\sqrt{\pi}s} H_{2,2}^{2,1}\left[\frac{1}{s^2}\left(\frac{1}{2\Omega_\ell p}\right)^p \bigg| \begin{array}{c}(1,2),(0,1)\\(\frac{1}{2},p),(1,1)\end{array}\right] = \frac{4}{\sqrt{(2\pi)^{p+1}}s} G_{3,p+1}^{p+1,2}\left[\frac{4}{s^2}\left(\frac{1}{2\Omega_\ell p}\right)^p \bigg| \begin{array}{c}1,\frac{1}{2},0\\\Xi_{(p)}^{(\frac{1}{2})},1\end{array}\right],$ where $G_{p,q}^{m,n}[\cdot]$ and $H_{p,q}^{m,n}[\cdot]$ represent the Meijer's G function [9, Eq. (8.2.1/1)] and Fox's H function [9, Eq. (8.3.1/1)], respectively. In addition, the coefficient $\Xi_{(n)}^{(x)}$ of the Meijer's G function is a set of coefficients such that it is defined as $\Xi_{(n)}^{(x)} \equiv \frac{x}{n}, \frac{x+1}{n}, \ldots, \frac{x+n-1}{n}$ with $x \in \mathbb{C}$ and $n \in \mathbb{N}$. |
| **Rayleigh** [1, Eq. (2.6)] $$p_{\mathcal{R}_\ell}(r) = \frac{2r}{\Omega_\ell} \exp\left(-\frac{r^2}{\Omega_\ell}\right)$$ defined over $r \in \mathbb{R}^+$, and where $\Omega_\ell$ is the average power (i.e., $\Omega_\ell \geq 0$). Note that Rayleigh fading distribution has unit AoF (that is, $AoF = 1$). | $\mathcal{M}_{\mathcal{R}_\ell^p}(s) = 2 H_{1,1}^{1,1}\left[\frac{1}{s^2\Omega_\ell^p} \bigg| \begin{array}{c}(1,2)\\(1,p)\end{array}\right] = \sqrt{\frac{2p}{(2\pi)^p}} G_{2,p}^{p,2}\left[\frac{4}{s^2(\Omega_\ell p)^p} \bigg| \begin{array}{c}\frac{1}{2},1\\\Xi_{(p)}^{(1)}\end{array}\right],$ $\frac{\partial}{\partial s}\mathcal{M}_{\mathcal{R}_\ell^p}(s) = \frac{4}{s} H_{2,2}^{2,1}\left[\frac{1}{s^2(\Omega_\ell p)^p} \bigg| \begin{array}{c}(1,2),(0,1)\\(1,p),(1,1)\end{array}\right] = \sqrt{\frac{8p}{(2\pi)^p}} \frac{1}{s} G_{3,p+1}^{p+1,2}\left[\frac{4}{s^2}\left(\frac{1}{2\Omega_\ell p}\right)^p \bigg| \begin{array}{c}1,\frac{1}{2},0\\\Xi_{(p)}^{(1)},1\end{array}\right],$ Rayleigh distribution typically agrees very well with experimental data for mobile systems where no line-of-sight (LOS) path exists between the transmitter and receiver antennas [1, Sec. 2.2.1.1]. |
| **Nakagami-$m$** [1, Eq. (2.20)] $$p_{\mathcal{R}_\ell}(r) = \frac{2}{\Gamma(m_\ell)}\left(\frac{m_\ell}{\Omega_\ell}\right)^{m_\ell} r^{2m_\ell-1} \exp\left(-\frac{m_\ell r^2}{\Omega_\ell}\right)$$ defined over $r \in \mathbb{R}^+$, where $\Omega_\ell$ is the average power, and where $m_\ell$ ($0.5 \leq m_\ell$) denotes the fading figure. Moreover, $\Gamma(\cdot)$ is the Gamma function [6, Sec. 8.31]. | $\mathcal{M}_{\mathcal{R}_\ell^p}(s) = \frac{2}{\Gamma(m_\ell)} H_{1,1}^{1,1}\left[\frac{1}{s^2}\left(\frac{m_\ell}{\Omega_\ell}\right)^p \bigg| \begin{array}{c}(1,2)\\(m_\ell,p)\end{array}\right] = \frac{\sqrt{2p^{2m_\ell-1}}}{\sqrt{(2\pi)^p}\Gamma(m_\ell)} G_{2,p}^{p,2}\left[\frac{1}{s^2}\left(\frac{m_\ell}{\Omega_\ell p}\right)^p \bigg| \begin{array}{c}\frac{1}{2},1\\\Xi_{(p)}^{(m_\ell)}\end{array}\right],$ $\frac{\partial}{\partial s}\mathcal{M}_{\mathcal{R}_\ell^p}(s) = \frac{4}{\Gamma(m_\ell)s} H_{2,2}^{2,1}\left[\frac{1}{s^2}\left(\frac{m_\ell}{\Omega_\ell}\right)^p \bigg| \begin{array}{c}(1,2),(0,1)\\(m_\ell,p),(1,1)\end{array}\right] = \frac{\sqrt{8p^{2m_\ell-1}}}{\sqrt{(2\pi)^p}\Gamma(m_\ell)s} G_{3,p+1}^{p+1,2}\left[\frac{4}{s^2}\left(\frac{m_\ell}{\Omega_\ell p}\right)^p \bigg| \begin{array}{c}1,\frac{1}{2},0\\\Xi_{(p)}^{(m_\ell)},1\end{array}\right],$ Note that the Nakagami-$m$ distribution spans via the $m$ parameter the widest range of amount of fading (AoF) among all the multipath distributions [1]. As such, Nakagami-$q$ (Hoyt) and Nakagami-$n$ (Rice) can also be closely approximated by Nakagami-$m$ distribution [1, Eq. (2.25)], [1, Eq. (2.26)]. |
| **Weibull** [1, Eq. (2.27)] $$p_{\mathcal{R}_\ell}(r) = 2\xi_\ell\left(\frac{\omega_\ell}{\Omega_\ell}\right)^{\xi_\ell} r^{2\xi_\ell-1} \exp\left(-\left(\frac{\omega_\ell}{\Omega_\ell}\right)^{\xi_\ell} r^{2\xi_\ell}\right)$$ defined over $r \in \mathbb{R}^+$, where $\omega_\ell = \Gamma(1 + 1/\xi_\ell)$ and where $\xi_\ell$ ($0 < \xi_\ell$) denotes the fading shaping factor. Moreover, $\Omega_\ell$ is the average power. | $\mathcal{M}_{\mathcal{R}_\ell^p}(s) = 2 H_{1,1}^{1,1}\left[\frac{1}{s^2}\left(\frac{\omega_\ell}{\Omega_{s\ell}}\right)^p \bigg| \begin{array}{c}(1,2)\\(1,\frac{p}{\xi_\ell})\end{array}\right] = \sqrt{\frac{2pkl}{(2\pi)^{2k+pl-2}}} G_{2k,pl}^{pl,2k}\left[\frac{\omega_\ell^{pk}(2k)^{2k}}{s^{2k}\Omega_{s\ell}^{pk}(pl)^{pl}} \bigg| \begin{array}{c}-\Xi_{(2k)}^{(-2k)}\\\Xi_{(pl)}^{(1)}\end{array}\right],$ $\frac{\partial}{\partial s}\mathcal{M}_{\mathcal{R}_\ell^p}(s) = \frac{4}{s} H_{2,2}^{2,1}\left[\frac{1}{s^2}\left(\frac{\omega_\ell}{\Omega_{s\ell}}\right)^p \bigg| \begin{array}{c}(1,2),(0,1)\\(1,\frac{p}{\xi_\ell}),(1,1)\end{array}\right] = \frac{\sqrt{8pk^3l}}{\sqrt{(2\pi)^{2k+pl-2}}s} G_{2k+1,pl+1}^{pl+1,2k}\left[\frac{\omega_\ell^{pk}(2k)^{2k}}{s^{2k}\Omega_{s\ell}^{pk}(pl)^{pl}} \bigg| \begin{array}{c}-\Xi_{(2k)}^{(-2k)},0\\\Xi_{(pl)}^{(1)},1\end{array}\right],$ where the Meijer's G representations are given for the rational value of the fading shaping factor $\xi_{s\ell}$ (that is, we let $\xi_{s\ell} = k/l$, where $k$, and $l$ are arbitrary positive integers.) through the medium of algebraic manipulations utilizing [9, Eq. (8.3.2.22)]. In addition, note that if $\mathcal{R}_\ell$ is a sample of a Weibull distribution with the fading shaping factor $\xi_\ell$, then $\mathcal{R}_\ell^\alpha$ is also a sample of a Weibull distribution with the fading shaping factor $\xi_\ell/\alpha$. |

TABLE II
UNIFIED MGFS OF SOME WELL-KNOWN FADING CHANNEL MODELS

| Envelope Distribution, i.e., $p_{\mathcal{R}_\ell}(r)$ | Unified MGF $\mathcal{M}_{\mathcal{R}_\ell^p}(s)$ and its derivative $\frac{\partial}{\partial s}\mathcal{M}_{\mathcal{R}_\ell^p}(s)$, where the exponent $p \in \{1,2\}$ |
|---|---|
| **Hyper Nakagami-$m$** [22, Eq. (1)] $p_{\mathcal{R}_\ell}(r) = \sum_{k=1}^{K} \frac{2\xi_{\ell k}}{\Gamma(m_{\ell k})} \left(\frac{m_{\ell k}}{\Omega_{\ell k}}\right)^{m_{\ell k}} r^{2m_{\ell k}-1} \exp\left(-\frac{m_{\ell k}}{\Omega_{\ell k}}r^2\right)$ defined over $r \in \mathbb{R}^+$, where $m_{\ell k}$ ($0.5 \leq m_{\ell k}$) is the fading figure, $\Omega_{\ell k}$ ($0 < \Omega_{\ell k}$) is the average power, and $\xi_{\ell k}$ ($0 < \xi_{\ell k}$) is the accruing factor, of the $k$th fading environment. | $\mathcal{M}_{\mathcal{R}_\ell^p}(s) = \sum_{k=1}^{K} \frac{2\xi_{\ell k}}{\Gamma(m_{\ell k})} H_{1,1}^{1,1}\left[\left(\frac{m_{\ell k}}{s^{\frac{2}{p}}\Omega_{\ell k}}\right)^p \bigg| \begin{array}{c}(1,2)\\(m_{\ell k},p)\end{array}\right] = \sum_{k=1}^{K} \frac{\sqrt{2p^{2m_\ell-1}}\xi_{\ell k}}{\sqrt{(2\pi)^p}\Gamma(m_{\ell k})} G_{2,p}^{p,2}\left[\left(\frac{2^{\frac{2}{p}}m_{\ell k}}{s^{\frac{2}{p}}\Omega_{\ell k}p}\right)^p \bigg| \begin{array}{c}\frac{1}{2},1\\ \Xi_{(p)}^{(m_{\ell k})}\end{array}\right],$  $\frac{\partial}{\partial s}\mathcal{M}_{\mathcal{R}_\ell^p}(s) = \sum_{k=1}^{K} \frac{4\xi_{\ell k}/s}{\Gamma(m_{\ell k})} H_{2,2}^{2,1}\left[\left(\frac{m_{\ell k}}{s^{\frac{2}{p}}\Omega_{\ell k}}\right)^p \bigg| \begin{array}{c}(1,2),(0,1)\\(m_{\ell k},p),(1,1)\end{array}\right] = \sum_{k=1}^{K} \frac{\sqrt{8p^{2m_{\ell k}-1}}\xi_{\ell k}}{\sqrt{(2\pi)^p}\Gamma(m_{\ell k})s} G_{3,p+1}^{p+1,2}\left[\left(\frac{2^{\frac{2}{p}}m_{\ell k}}{s^{\frac{2}{p}}\Omega_{\ell k}p}\right)^p \bigg| \begin{array}{c}1,\frac{1}{2},0\\ \Xi_{(p)}^{(m_{\ell k})},1\end{array}\right],$ where $\Gamma(\cdot)$ is the Gamma function [11, Eq. (6.1.1)]. In addition, It may be useful to notice that the sum of the accruing probabilities $\xi_{\ell k}$, $k \in \{1,2,\ldots,K\}$ of $K$ possible fading environments is unit such that $\sum_{k=1}^{K}\xi_{\ell k}=1$. |
| **Nakagami-$q$ (Hoyt)** [1, Eq. (2.10)] $p_{\mathcal{R}_\ell}(r) = \frac{(1+q_\ell^2)r}{q_\ell\Omega_\ell}\exp\left(-\frac{(1+q_\ell^2)^2 r^2}{4q_\ell^2\Omega_\ell}\right)I_0\left(\frac{1-q_\ell^4}{4q_\ell^2\Omega_\ell}r^2\right)$ defined over $r \in \mathbb{R}^+$, where $q_\ell$ ($0 < q_\ell < 1$) is the Nakagami-$q$ fading parameter (that is, it is defined as ratio of the powers of the received signal's in-phase and quadrature with different standard deviations), and where $\Omega_\ell$ ($0 < \Omega_\ell$) is the average power. In addition, $I_0(\cdot)$ is the zeroth order modified Bessel function of the first kind [11, Eq. (9.6.20)]. | $\mathcal{M}_{\mathcal{R}_\ell^p}(s) = \frac{1+q_\ell^2}{q_\ell\Phi_\ell}\sum_{k=0}^{\infty}\frac{\Psi_k}{(2k)!}H_{1,1}^{1,1}\left[\frac{1}{s^2}\left(\frac{\Phi_\ell}{\Omega_\ell}\right)^p\bigg|\begin{array}{c}(1,2)\\(2k+1,p)\end{array}\right] = \frac{1+q_\ell^2}{q_\ell\Phi_\ell}\sum_{k=0}^{\infty}\frac{p^{2k+1}\Psi_k}{\sqrt{2p(2\pi)^p}(2k)!}G_{2,p}^{p,2}\left[\frac{1}{s^2}\left(\frac{\Phi_\ell}{\Omega_\ell p}\right)^p\bigg|\begin{array}{c}1,\frac{1}{2}\\ \Xi_{(p)}^{(2k+1)}\end{array}\right],$  $\frac{\partial}{\partial s}\mathcal{M}_{\mathcal{R}_\ell^p}(s) = \frac{1+q_\ell^2}{sq_\ell\Phi_\ell}\sum_{k=0}^{\infty}\frac{2\Psi_k}{(2k)!}H_{2,2}^{2,1}\left[\frac{1}{s^2}\left(\frac{\Phi_\ell}{\Omega_\ell}\right)^p\bigg|\begin{array}{c}(1,2),(0,1)\\(2k+1,p),(1,1)\end{array}\right] = \frac{1+q_\ell^2}{sq_\ell\Phi_\ell}\sum_{k=0}^{\infty}\frac{2p^{2k+1}\Psi_k}{\sqrt{2p(2\pi)^p}(2k)!}G_{3,p+1}^{p+1,2}\left[\frac{1}{s^2}\left(\frac{\Phi_\ell}{\Omega_\ell p}\right)^p\bigg|\begin{array}{c}1,\frac{1}{2},0\\ \Xi_{(p)}^{(2k+1)},1\end{array}\right],$ where $\Phi_\ell$ is defined as $\Phi_\ell = 0.25(1+q^2)^2/q^2$, and $\Psi_k$ is given by $\Psi_k(q) = \frac{(2k)!}{(k!)^2 2^{2k}}\left((1-q^2)/(1+q^2)\right)^{2k}$, where $k \in \mathbb{N}$. It may be useful to notice that the series expression of the unified MGF for the Nakagami-$q$ (Hoyt) is converging very fast such that 10 summation terms is generally enough. |
| **Nakagami-$n$ (Rice)** [1, Eq. (2.15)] $p_{\mathcal{R}_\ell}(r) = \frac{2(1+n_\ell^2)e^{-n_\ell^2}r}{\Omega_\ell}e^{-\frac{(1+q_\ell^2)}{\Omega_\ell}r^2}I_0\left(2n_\ell\sqrt{\frac{1+n_\ell^2}{\Omega_\ell}}r^2\right)$ defined over $r \in \mathbb{R}^+$, where $n_\ell$ ($0 < n_\ell$) and $\Omega_\ell$ ($0 < \Omega_\ell$) are the LOS figure and average power, respectively. | $\mathcal{M}_{\mathcal{R}_\ell^p}(s) = 2\sum_{k=0}^{\infty}\frac{\mathcal{Z}_{\ell k}}{k!}H_{1,1}^{1,1}\left[\frac{1}{s^2}\left(\frac{1+n_\ell^2}{\Omega_\ell}\right)^p\bigg|\begin{array}{c}(1,2)\\(k+1,p)\end{array}\right] = \theta_p\sum_{k=0}^{\infty}\frac{p^k\mathcal{Z}_{\ell k}}{k!}G_{2,p}^{p,2}\left[\frac{1}{s^2}\left(\frac{1+n_\ell^2}{\Omega_\ell p}\right)^p\bigg|\begin{array}{c}1,\frac{1}{2}\\ \Xi_{(p)}^{(k+1)}\end{array}\right],$  $\frac{\partial}{\partial s}\mathcal{M}_{\mathcal{R}_\ell^p}(s) = \frac{4}{s}\sum_{k=0}^{\infty}\frac{\mathcal{Z}_{\ell k}}{k!}H_{2,2}^{2,1}\left[\frac{1}{s^2}\left(\frac{1+n_\ell^2}{\Omega_\ell}\right)^p\bigg|\begin{array}{c}(1,2),(0,1)\\(k+1,p),(1,1)\end{array}\right] = \frac{2\theta_p}{s}\sum_{k=0}^{\infty}\frac{p^k\mathcal{Z}_{\ell k}}{k!}G_{3,p+1}^{p+1,2}\left[\frac{1}{s^2}\left(\frac{1+n_\ell^2}{\Omega_\ell p}\right)^p\bigg|\begin{array}{c}1,\frac{1}{2},0\\ \Xi_{(p)}^{(k+1)},1\end{array}\right],$ where $\mathcal{Z}_{\ell k} = \eta^{2k}\exp(-n_\ell^2)/k!$ and the coefficient $\theta_p = \sqrt{2p}/\sqrt{(2\pi)^p}$. In addition, the LOS figure i.e. $n_\ell$ is related to the Rician $K_\ell$ factor by $K_\ell = n_\ell^2$ which corresponds to the ratio of the power of the LOS (specular) component to the average power of the scattered component. |
| **K-Distribution** [1, Eq. (2.15)] $p_{\mathcal{R}_\ell}(r) = \frac{4\left(\frac{m_{s\ell}}{\Omega_{s\ell}}\right)^{\frac{m_{s\ell}+1}{2}}}{\Gamma(m_{s\ell})}r^{m_{s\ell}}K_{m_{s\ell}-1}\left(2\sqrt{\frac{m_{s\ell}r^2}{\Omega_{s\ell}}}\right)$ defined over $r \in \mathbb{R}^+$, where $m_{s\ell}$ ($\frac{1}{2} \leq m_{s\ell}$) denotes the shadowing severity, and $\Omega_{s\ell}$ ($0 < \Omega_{s\ell}$) represents the average power. Moreover, $K_n(\cdot)$ is the $n$th order modified Bessel function of the second kind [11, Eq. (9.6.24)]. | $\mathcal{M}_{\mathcal{R}_\ell^p}(s) = \frac{2}{\Gamma(m_{s\ell})}H_{1,2}^{2,1}\left[\frac{1}{s^2}\left(\frac{m_{s\ell}}{\Omega_{s\ell}}\right)^p\bigg|\begin{array}{c}(1,2)\\(1,p),(m_{s\ell},p)\end{array}\right] = \frac{2\sqrt{\pi}p^{m_{s\ell}}}{(2\pi)^p\Gamma(m_{s\ell})}G_{2,p}^{p,2}\left[\frac{4}{s^2}\left(\frac{m_{s\ell}}{\Omega_{s\ell}p^2}\right)^p\bigg|\begin{array}{c}1,\frac{1}{2}\\ \Xi_{(p)}^{(m_{s\ell})},\Xi_{(p)}^{(1)}\end{array}\right],$  $\frac{\partial}{\partial s}\mathcal{M}_{\mathcal{R}_\ell^p}(s) = \frac{2}{\Gamma(m_{s\ell})}H_{1,2}^{2,1}\left[\frac{1}{s^2}\left(\frac{m_{s\ell}}{\Omega_{s\ell}}\right)^p\bigg|\begin{array}{c}(1,2)\\(1,p),(m_{s\ell},p)\end{array}\right] = \frac{2\sqrt{\pi}p^{m_{s\ell}}}{(2\pi)^p\Gamma(m_{s\ell})}G_{2,p}^{p,2}\left[\frac{4}{s^2}\left(\frac{m_{s\ell}}{\Omega_{s\ell}p^2}\right)^p\bigg|\begin{array}{c}1,\frac{1}{2}\\ \Xi_{(p)}^{(m_{s\ell})},\Xi_{(p)}^{(1)}\end{array}\right],$ It may be useful to notice that the shadowing effect in the channel disappears when $m_{s\ell}$ approaches to infinity ($m_{s\ell} \to \infty$) such that the worst shadowing occurs when $m_{s\ell} = \frac{1}{2}$. |

TABLE III
UNIFIED MGFs OF SOME WELL-KNOWN FADING CHANNEL MODELS

| **Envelope Distribution, i.e., $p_{\mathcal{R}_\ell}(r)$** | **Unified MGF $\mathcal{M}_{\mathcal{R}_\ell^p}(s)$ and its derivative $\frac{\partial}{\partial s}\mathcal{M}_{\mathcal{R}_\ell^p}(s)$, where the exponent $p \in \{1,2\}$** |
|---|---|
| **Generalized-K** [23, Eq. (5)] $$p_{\mathcal{R}_\ell}(r) = \frac{4\left(\frac{m_{s\ell}m_\ell}{\Omega_{s\ell}}\right)^{\frac{\phi_\ell}{2}}}{\Gamma(m_{s\ell})\Gamma(m_\ell)} r^{\phi_\ell-1} K_{\psi_\ell}\left(2\sqrt{\frac{m_{s\ell}m_\ell r^2}{\Omega_{s\ell}}}\right)$$ defined over $r \in \mathbb{R}^+$, where $\phi_\ell = m_{s\ell}+m_\ell$ and $\psi_\ell = m_{s\ell}-m_\ell$. Moreover, $m_\ell$ ($0.5 \leq m_\ell$) and $m_{s\ell}$ ($0.5 \leq m_{s\ell}$) represent the fading figure (diversity severity / order) and the shadowing severity, respectively. $\Omega_{s\ell}$ ($0 < \Omega_{s\ell}$) represents the average power. | $$\mathcal{M}_{\mathcal{R}_\ell^p}(s) = \frac{2}{\mathcal{G}_\ell} \mathrm{H}_{1,2}^{2,1}\left[\frac{1}{s^2}\left(\frac{m_{s\ell}m_\ell}{\Omega_{s\ell}}\right)^p \middle| \begin{array}{c}(1,2)\\(m_\ell,p),(m_{s\ell},p)\end{array}\right] = \frac{\sqrt{2}\,p^{m_{s\ell}+m_\ell-1}}{\sqrt{(2\pi)^{2p-1}}\mathcal{G}_\ell} \mathrm{G}_{2,2p}^{2p,2}\left[\frac{4}{s^2}\left(\frac{m_{s\ell}m_\ell}{\Omega_{s\ell}p^2}\right)^p \middle| \begin{array}{c}1,\frac{1}{2}\\ \Xi_{(p)}^{(m_{s\ell})}, \Xi_{(p)}^{(m_\ell)}\end{array}\right],$$ $$\frac{\partial}{\partial s}\mathcal{M}_{\mathcal{R}_\ell^p}(s) = \frac{4}{\mathcal{G}_\ell s} \mathrm{H}_{2,3}^{3,1}\left[\frac{1}{s^2}\left(\frac{m_{s\ell}m_\ell}{\Omega_{s\ell}}\right)^p \middle| \begin{array}{c}(1,2),(0,1)\\(m_\ell,p),(m_{s\ell},p),(1,1)\end{array}\right] = \frac{2\sqrt{2}\,p^{m_{s\ell}+m_\ell-1}}{\sqrt{(2\pi)^{2p-1}}\mathcal{G}_\ell\, s} \mathrm{G}_{3,2p+1}^{2p+1,2}\left[\frac{4}{s^2}\left(\frac{m_{s\ell}m_\ell}{\Omega_{s\ell}p^2}\right)^p \middle| \begin{array}{c}1,\frac{1}{2},0\\ \Xi_{(p)}^{(m_{s\ell})}, \Xi_{(p)}^{(m_\ell)}\end{array}\right],$$ where $\mathcal{G}_\ell = \Gamma(m_{s\ell})\Gamma(m_\ell)$. It may be useful to notice that the shadowing effect in the channel disappears and generalized-K distribution turns into Nakagami-$m$ when $m_{s\ell}$ approaches to infinity ($m_{s\ell} \to \infty$) such that the worst shadowing occurs when $m_{s\ell}=\frac{1}{2}$. |
| **Composite Nakagami / Lognormal** [1, Eq. (2.57)] $$p_{\mathcal{R}_\ell}(r) = \frac{2r^{2m_\ell-1}}{\Gamma(m_\ell)} \int_{-\infty}^{\infty} \left(\frac{m_\ell}{\mathcal{G}_\ell(u)}\right)^{m_\ell} e^{-\left(\frac{m_\ell r^2}{\mathcal{G}_\ell(u)}+u^2\right)} du$$ defined over $r \in \mathbb{R}^+$, where $\mathcal{G}_\ell(u) = 10^{(\sqrt{2}\sigma_\ell u + \mu_\ell)/10}$, and where $\mu_\ell$(dB) and $\sigma_\ell$(dB) are the mean and the standard deviation of channel shadowing. Moreover, $m_\ell$ ($0.5 \leq m_\ell$) is the fading figure (diversity order), and $\Omega_\ell$ ($0 < \Omega_\ell$) represents the average power. | $$\mathcal{M}_{\mathcal{R}_\ell^p}(s) = \frac{1}{\pi}\sum_{n=1}^{N_p} \frac{\mathrm{H}_{x_n}}{\Gamma(m_\ell)} \mathrm{H}_{2,1}^{1,2}\left[\frac{4}{s^2}\left(\frac{m_\ell}{\mathcal{G}_\ell(x_n)}\right)^p \middle| \begin{array}{c}(1,1),(\frac{1}{2},1)\\(m,p)\end{array}\right] = \frac{2p^{m_\ell-\frac{1}{2}}}{(2\pi)^{\frac{p+1}{2}}}\sum_{n=1}^{N_p} \frac{\mathrm{H}_{x_n}}{\Gamma(m_\ell)} \mathrm{G}_{2,p}^{p,2}\left[\frac{4}{s^2}\left(\frac{m_\ell}{\mathcal{G}_\ell(x_n)p}\right)^p \middle| \begin{array}{c}1,\frac{1}{2}\\ \Xi_{(p)}^{(m_\ell)}\end{array}\right],$$ $$\frac{\partial}{\partial s}\mathcal{M}_{\mathcal{R}_\ell^p}(s) = \frac{1}{\pi s}\sum_{n=1}^{N_p} \frac{\mathrm{H}_{x_n}}{\Gamma(m_\ell)} \mathrm{H}_{3,2}^{2,2}\left[\frac{4}{s^2}\left(\frac{m_\ell}{\mathcal{G}_\ell(x_n)}\right)^p \middle| \begin{array}{c}(1,1),(\frac{1}{2},1),(0,1)\\(m,p),(1,1)\end{array}\right] = \frac{2p^{m_\ell-\frac{1}{2}}}{s(2\pi)^{\frac{p+1}{2}}}\sum_{n=1}^{N_p} \frac{\mathrm{H}_{x_n}}{\Gamma(m_\ell)} \mathrm{G}_{3,p+1}^{p+1,2}\left[\frac{4}{s^2}\left(\frac{m_\ell}{\mathcal{G}_\ell(x_n)p}\right)^p \middle| \begin{array}{c}1,\frac{1}{2},0\\ \Xi_{(p)}^{(m_\ell)},1\end{array}\right],$$ where, for $n \in \{1,2,\ldots,N_p\}$, $\{H_{x_n}\}$ and $\{x_n\}$ are the weight factors and the zeros (abscissas) of the $N_p$-order Hermite polynomial [11, Table 25.10]. |
| **Composite Nakagami / Weibull** [24, Eq. (4)] $$p_{\mathcal{R}_\ell}(r) = \frac{2}{\Gamma(m_{s\ell})r} \mathrm{H}_{0,2}^{2,0}\left[\frac{m_{s\ell}\omega_\ell}{\Omega_\ell}r^2 \middle| \begin{array}{c}---\\(m_{s\ell}),(1,\frac{1}{\xi_\ell})\end{array}\right]$$ defined over $r \in \mathbb{R}^+$, where $\Omega_\ell$ ($0 < \Omega_\ell$) is the average power and $\xi_\ell$ ($0 < \xi_\ell$) denotes the Weibull (fading shaping) factor chosen to yield a best fit to measurement results. In addition, $\omega_\ell = \Gamma(1+1/\xi_\ell)$ and $m_{s\ell}$ ($0.5 \leq m_{s\ell}$) is the shadowing severity. | $$\mathcal{M}_{\mathcal{R}_\ell^p}(s) = \frac{2}{\Gamma(m_{s\ell})} \mathrm{H}_{1,2}^{2,1}\left[\frac{1}{s^2}\left(\frac{m_{s\ell}\omega_\ell}{\Omega_\ell}\right)^p \middle| \begin{array}{c}(1,2)\\(m_{s\ell},p),(1,\frac{p}{\xi_\ell})\end{array}\right],$$ $$\frac{\partial}{\partial s}\mathcal{M}_{\mathcal{R}_\ell^p}(s) = \frac{4}{s\,\Gamma(m_{s\ell})} \mathrm{H}_{2,3}^{3,1}\left[\frac{1}{s^2}\left(\frac{m_{s\ell}\omega_\ell}{\Omega_{s\ell}}\right)^p \middle| \begin{array}{c}(1,2),(0,1)\\(m_{s\ell},p),(1,\frac{p}{\xi_\ell}),(1,1)\end{array}\right],$$ Note that Composite Nakagami / Lognormal distribution is the special case of Gamma-shadowed GNM distribution so the Meijer's G representation of the composite Nakagami / Lognormal distribution can be readily obtained by means of substituting $m_\ell=1$ and $\Omega_{s\ell}=\Omega_\ell$ into both (26) and (28). |
| **Fox's H distribution** [25, Eq. (3.1)], [26] $$p_{\mathcal{R}_\ell}(r) = \mathcal{K}_\ell \mathrm{H}_{p,q}^{m,n}\left[\mathcal{G}_\ell\, r \middle| \begin{array}{c}(a_1,\alpha_1),(a_2,\alpha_2),\ldots,(a_n,\alpha_n)\\(b_1,\beta_1),(b_2,\beta_2),\ldots,(b_m,\beta_m)\end{array}\right]$$ defined over $r \in \mathbb{R}^+$, and where $\mathcal{K}_\ell \in \mathbb{R}$ and $\mathcal{G}_\ell \in \mathbb{R}$ are such two numbers that $\int_0^\infty p_{\mathcal{R}_\ell}(r)\,dr = 1$. | $$\mathcal{M}_{\mathcal{R}_\ell^p}(s) = \frac{\mathcal{K}_\ell}{ps^{\frac{1}{p}}} \mathrm{H}_{p+1,q}^{m,n+1}\left[\frac{\mathcal{G}_\ell}{s^{\frac{1}{p}}} \middle| \begin{array}{c}(1-\frac{1}{p},\frac{1}{p}),(a_1,\alpha_1),(a_2,\alpha_2),\ldots,(a_n,\alpha_n)\\(b_1,\beta_1),(b_2,\beta_2),\ldots,(b_m,\beta_m)\end{array}\right],$$ $$\frac{\partial}{\partial s}\mathcal{M}_{\mathcal{R}_\ell^p}(s) = -\frac{\mathcal{K}_\ell}{p\,s^{\frac{p+1}{p}}} \mathrm{H}_{p+1,q}^{m,n+1}\left[\frac{\mathcal{G}_\ell}{s^{\frac{1}{p}}} \middle| \begin{array}{c}(-\frac{1}{p},\frac{1}{p}),(a_1,\alpha_1),(a_2,\alpha_2),\ldots,(a_n,\alpha_n)\\(b_1,\beta_1),(b_2,\beta_2),\ldots,(b_m,\beta_m)\end{array}\right],$$ |